\documentclass[twoside]{article}
\usepackage{qic,epsfig}
\usepackage{amsmath}
\usepackage{epstopdf} 

\begin{document}
\setlength{\textheight}{8.0truein}    

\runninghead{Arithmetic Circuits for Multilevel Qudits  Based on Quantum Fourier Transform}
{A. Pavlidis and E. Floratos}

\normalsize\textlineskip
\thispagestyle{empty}
\setcounter{page}{1}


\vspace*{0.88truein}

\alphfootnote

\fpage{1} 

\centerline{\bf
ARITHMETIC CIRCUITS FOR MULTILEVEL QUDITS}
\vspace*{0.035truein}
\centerline{\bf BASED ON QUANTUM FOURIER TRANSFORM }
\vspace*{0.37truein}
\centerline{\footnotesize
ARCHIMEDES PAVLIDIS}
\vspace*{0.015truein}
\centerline{\footnotesize\it Department of Informatics, University of Piraeus}
\baselineskip=10pt
\centerline{\footnotesize\it 80, Karaoli \& Dimitriou str., GR 185 34, Piraeus, Greece}
\centerline{\footnotesize\it Department of Informatics and Telecommunications, National and Kapodistrian University of Athens}
\baselineskip=10pt
\centerline{\footnotesize\it Panepistimiopolis, Ilissia, GR 157 84, Athens, Greece}
\baselineskip=10pt
\centerline{\footnotesize\it e-mail: adp@unipi.gr}
\vspace*{10pt}
\centerline{\footnotesize 
EMMANUEL FLORATOS}
\vspace*{0.015truein}
\centerline{\footnotesize\it Department of Physics, National and Kapodistrian University of Athens}
\baselineskip=10pt
\centerline{\footnotesize\it Panepistimiopolis, Ilissia, GR 157 84, Athens, Greece}
\baselineskip=10pt
\centerline{\footnotesize\it Institute of Nuclear and Particle  Physics, N.C.S.R. Demokritos}
\baselineskip=10pt
\centerline{\footnotesize\it 27, Neapoleos str., Agia Paraskevi, GR 153 41, Athens, Greece}
\baselineskip=10pt
\centerline{\footnotesize\it e-mail: mflorato@phys.uoa.gr}
\vspace*{0.225truein}

\vspace*{0.21truein}

\abstracts{
We present some basic integer arithmetic quantum circuits, such as adders and multipliers-accumulators of various forms, as well as  diagonal operators,  which operate on multilevel qudits. The integers to be processed are represented in an alternative basis after they have been Fourier transformed. Several arithmetic circuits operating on Fourier transformed integers have appeared in the literature for two level qubits. Here we extend these techniques on multilevel qudits, as they may offer some advantages relative to qubits implementations. The arithmetic circuits presented can be used as basic building blocks for higher level algorithms such as quantum phase estimation, quantum simulation, quantum optimization etc., but they can also be used in the implementation of a quantum fractional Fourier transform as it is shown in a companion work presented separately.  
}{}{}

\vspace*{10pt}

\vspace*{3pt}

\vspace*{1pt}\textlineskip    
\section{Introduction}\label{sec:Introduction}     
  
\noindent 
The common representation of the elementary quantum information is the qubit, where its state is a superposition $a|0\rangle+b|1\rangle$ which belongs to a two-dimensional Hilbert space with two basis states $|0\rangle$ and $|1\rangle$ known as the computational basis. A quantum computer is a finite dimensional quantum system composed of a  qubits collection, performing various unitary operations on the qubits (quantum gates) and quantum measurements. Accordingly, there is a correspondence between a qubit and a classical bit, in the sense  that the basis states of a qubit follow the binary logic. We can extend this correspondence to multivalued logic instead of two values only by enlarging the dimension of the elementary Hilbert space used. The \emph{qudit} is a generalization of the qubit to a larger Hilbert space of dimension $d>2$. The state of a qudit is a superposition $a_{0}|0\rangle+a_{1}|1\rangle+\cdots + a_{d-1}|d-1\rangle$, where $|0\rangle, |1\rangle, \ldots  |d-1\rangle$ are the computational basis states. \emph{Qutrit} is a special name for the case  $d=3$, while \emph{ququart} corresponds to $d=4$. In many cases, the employment of a multivalued quantum logic is more natural. E.g. in ion traps we could exploit more than two energy levels. Multiple laser beams could be used to manipulate the transitions between these levels \cite{Mut:2000}. 

Working with qudits instead of qubits may offer some advantages. The required number of qudits is smaller by a factor $\log _{2}d$ than the corresponding number of qubits for the same dimension a quantum computer has to explore. E.g. the dimensions of a composite system of $n$ qubits  is $2^{n}$, while the same dimension  can be reached with only $\log_{d}2^{n}=\log_{2}2^{n} / \log_{2}d=n/log_{2}d$ qudits. Such as reduction of the required number of physical carriers of quantum information is advantageous, considering the difficulty of reliably controlling a large number of  carriers. Also, when fewer quantum information carriers are used, a decrease in the overall decoherence is expected and this fact favors the scalability issues \cite{Mut:2000,Mut:2002}. 

Another advantage, which is also related to the adverse effect of decoherence, is that fewer multilevel qudit gates are required to construct a quantum circuit implementing a given unitary operation compared to the case of using two-level gates \cite{Mut:2000,Lan:2009}.  
Fewer gates reduce the number of steps needed to complete the circuit operation (circuit depth), and consequently less errors are accumulated during the overall operation of the circuit. Even so, protection of quantum information against environmental interaction is inevitable. Quantum error correcting codes and fault tolerant gate constructions to combat decoherence on multilevel qudits have been proposed and they are similar to the ones used for the qubit case \cite{Got:1999,Ket:2006,Cam:2014}. 

At a higher level, generalizations of known quantum algorithms and circuits using $d$-level qudits may offer improvements with respect to their qubits implementation counterparts. E.g., quantum phase estimation, which is the core part of Shor's algorithm \cite{Shor:1994}	 and also it is used in quantum simulation \cite{Asp:2005}, is improved in terms of success probability when multilevel qudits are incorporated \cite{Par:2011}. Multiple-valued version of Deutsch-Josza algorithm  has been reported in \cite{Fan:2007} while an implementation proposal for five level superconducting qudit appeared in \cite{Kik:2015}. Qudits version for Grover's algorithm \cite{Grover:1996} has been reported in \cite{Iva:2012}. The high dimensional Deutsch-Josza algorithm may find applications in image processing, while the high dimensional Grover's algorithm offers a trade-off between space and time. 

An assortment of quantum gates operating on qudits have been proposed or experimentally realized on various technologies. Single and two qudit $d$-level gates proposed in \cite{Mut:2000,Di:2013} for the ion trap technology. Single qudit gates for $d=5$ implemented in superconducting technology and used to emulate spins of 1/2, 1 and 3/2 in \cite{Nee:2009}. Proposals for single and two qudit gates based on superconducting technology appeared in \cite{Str:2011}. Single qudit gates based on optical technology reported in \cite{Bab:2017}. Three dimensional entanglement between photons observed in \cite{Mal:2016}.

In this work we present some quantum arithmetic circuits operating on $d$-level qudits by extending results given in prior works \cite{Dra:1998,Bea:2003,Pav:2014}. These circuits exploit the quantum Fourier transform and various single qudit and two qudit rotation gates to perform the desired calculations. Processing in the Fourier domain may offer some advantages related to speed \cite{Pav:2014} and robustness to decoherence \cite{Bar:1996,Nam:2015}. Among the proposed circuits are various versions of adders (adder with constant, generic adder, adder with constant controlled by single qudit) and multipliers  (multiplier with constant and accumulator, multiplier with constant). Such circuits are useful in many quantum algorithms, e.g. quantum phase estimation, quantum simulation. 

The increased interest in quantum information processing exploiting $d$-level qudits, both in theoretical and experimental aspects, was one of the stimulation for this work. 
However, the main motivation was the particular application targeted by the quantum circuits presented in this manuscript, which is a new definition of the fractional Fourier transform. Unitary operations on high dimensional $d$-level qudits fit more naturally for this specific application because the proposed fractional Fourier transform operates on a Hilbert space of dimension $d^n$, where $d \neq 2$ is a prime. The development of the \emph{quantum fractional Fourier transform} and its implementation on qudits is presented in a separate work \cite{Flo:2017}.

The rest of the paper is organized as follows: A short background about design and synthesis of qudits quantum circuits is given in section \ref{sec:Background}. The elementary and basic qudit gates used in the proposed designs are given in section \ref{sec:Elementary}. The quantum Fourier transform definition and its circuit for $q$ qudits of $d$ levels is presented in section \ref{sec:QFT}. Section \ref{sec:Arithmetic} introduces integer arithmetic circuits like adder with constant, adder of two integers, controlled adder with constant, multiplier with constant and accumulator, and multiplier with constant. All of the arithmetic units accept one of their operands after it has been Fourier transformed.  In section \ref{sec:DiagonalOperators} a method to implement a diagonal operator on $q$ qubits is analyzed where the diagonal elements are some powers of roots of unity. A quantum multiplier of two integers is introduced  and then a quantum squarer is built upon this multiplier. It is demonstrated how to introduce relative phases between the basis states of superposition which depend quadratically on the index of the basis state. It can be generalized for a function that is polynomial in the states index. This operation is a necessary part of  the quantum fractional Fourier transform presented in the companion article and also it may find other applications, such as quantum simulation algorithms and Grover's search algorithm. Appendix A gives the decomposition of a three qudits rotation gate introduced in section \ref{sec:DiagonalOperators}. Complexity analysis in terms of quantum cost, depth and width is reported in section \ref{sec:Complexity}. In Appendix B we discuss how it is possible to use a discrete library of components to approximate the proposed designs and the impact to cost and depth. A discrete library of gates is necessary if fault tolerance is to be incorporated. Finally, we conclude in section \ref{sec:Conclusion}.

\section{Background and related work}\label{sec:Background}

\noindent
The construction of a complex quantum circuit operating on multilevel qudits is based on the selection of a set of elementary qudit gates and their interconnection so as to achieve the target operation. A multilevel gate operates on a single qudit, on two qudits or more qudits. A single $d$-level qudit gate is represented by a unitary matrix $U$ of dimensions $d\times d$. It transforms an initial qudit state $|\psi  \rangle _{in}$ to $|\psi  \rangle _{out}=U|\psi  \rangle _{in}$. As an example consider the  general superposition qutrit state $|\psi  \rangle _{in}=a|0\rangle+b|1\rangle+c|2\rangle$. The application of the gate 
\[ 
U=
\begin{bmatrix}
0 & 0 & 1 \\
1 & 0 & 0 \\
0 & 1 & 0
\end{bmatrix}
=|1\rangle \langle 0| + |2\rangle \langle 1|  + |0\rangle \langle 2|
\]

\noindent on this state results to the state $|\psi  \rangle _{out}= c|0\rangle+a|1\rangle+b|2\rangle$.

Two qudit gates operate on states of two qudits which are $d^2$ dimensional, so their representing unitary matrices have dimensions of $d^2 \times d^2$. Two single qudit gates $V_{1}$ and $V_{2}$ operating on two different qudits can be seen as a two qudit gate which is their tensor product $U=V_{1}\otimes V_{2}$. However, not every two qudit gate can be decomposed as a tensor product of two single qudit gates, in which case we have an \emph{entangling} gate.  

Consecutive application of single, two or more qudit gates to a collection of $q$ qudits results in a quantum circuit which is represented by a unitary matrix of dimensions $d^q \times d^q$. The design of a quantum circuit is the procedure of interconnecting various elementary gates so as to fulfill the given specifications. These specifications are given in the form of a unitary matrix or the relationship between the desired input-output state relation in the computational basis.

It is proven that single qudit gates  and a two qudit gate alone are adequate to form a \emph{universal} set of gates, provided that the two qudit gate is an entangling gate \cite{Bry:2002}. A universal set of gates can be used to approximate any target  quantum circuit with arbitrary precision. Various sets of qudit gates (\emph{gate libraries}) and methods to exploit them to build more complex unitaries have been introduced in the literature. The library used in  \cite{Mut:2000} consists of single and two qudit gates with continuous parameters and the synthesis method is based on spectral decomposition of the target unitary matrix. Cosine-Sine decomposition is another method used in  \cite{Kha:2006}. A discrete set of single qudit gates and a single two qudit gate is used in  \cite{Bre:2005}  to synthesize the large unitary matrix using QR decomposition. In \cite{Di:2013} a different two qudit gate and a set of single qudit gates is used along with quantum Shannon decomposition to synthesize the target unitary. The previous methods and results are similar to the two-level qubits synthesis cases.  It is proven that the cost of the resulting circuit in terms of two qudit gates is  upper bounded by $O(d^{2n})$ where $n$ is the qudits number \cite{Bul:2005}. Thus, these automated methods are suitable only for small quantum circuits due to the exponential cost increase. 

When the target circuit is an arithmetic or logic block where its unitary matrix is a permutation matrix consisting of 0 and 1 elements, then multiple-valued reversible synthesis methods could be applied. These methods are extension of the binary reversible logic case and may be applied to a specific value of $d$, e.g. \cite{Per:2004} ($d=3$), or applied to any value of $d$ \cite{Mil:2004,Den:2004}. Similarly to the quantum synthesis case, these algorithms are not suitable for large circuits.

As many algorithms widely use quantum arithmetic blocks like adders or multipliers recurrently, it is crucial to have available efficient arithmetic and logic blocks. Ad hoc design of such blocks usually offers better results compared to the automated synthesis methods. One can exploit the iterative and regular structure of these arithmetic blocks or extend known classical designs to the quantum case. A diversity of ad hoc designed quantum arithmetic and logic circuits for two-level qubits can be found in the literature \cite{Ved:1996,Dra:1998,Cuc:2005,VanM:2005,Dra:2006,Kho:2011,Cho:2012,Pav:2014}, but few (usually adders) are known for multilevel qudits and also they are mostly designed for a specific value of $d$. In contrast, the proposed designs are parametrized for any value of $d$.

One of the first \emph{ternary} ($d=3$) quantum adder for 3-inputs only appeared in \cite{Mil:2004} as an application example of the proposed synthesis method. Ternary quantum adders/subtractors ad hoc designed for any number of inputs is given in \cite{Kha:2007}. In \cite{Sat:2007} a ternary extension of the well known VBE ripple-carry adder \cite{Ved:1996} is reported. \emph{Quaternary} ($d=4$) comparators proposed in \cite{Kha:2008}. Improved designs of ternary ripple carry and carry look-ahead adders along with modifications that lead to subtractors and comparators are given in  \cite{Boch:2016}. The previous ternary ripple carry adder is a modification of the CDKM binary quantum adder appeared in  \cite{Cuc:2005}  and it has also a depth of $O(n)$ using one ancilla qutrit. Similarly, the previous ternary carry look-ahead quantum adder is an extension of the DKRS binary quantum adder appeared in \cite{Dra:2006} and thus it offers a depth $O(log(n))$ using $O(n)$ ancilla qutrits. 

Several of the previous multilevel qudits (usually qutrits) designs are modifications of binary quantum adders. The gates libraries used are differentiated among each design. This is justifiable as the implementation technology for multilevel qudits is far apart to be considered matured. However, gates of one library can be expressed as gates of another one, provided that the libraries are universal. 

Diagonal operator circuits on qubits or qudits don't change the absolute value of the amplitudes of a superposition, but rearrange their relative phases. Such circuits are useful in quantum algorithms \cite{Zal:1998,Kas:2008,Wel:2014} like quantum optimization, quantum simulation, Grover's search etc.  Synthesis methods for diagonal unitary matrices of two-level qubits  have been developed  for diagonals of special structure \cite{Hog:1999,Wel:2014,Wel:2016} or for any diagonal \cite{Bul:2004}. Recently, diagonal synthesis for multilevel qudits reported in \cite{Bee:2016}. In general, the synthesis cost is related to the dimensionality of the Hilbert space covered (that is exponential in the number of the qubits or qudits) and the number of the distinct phases of the diagonal. In this work, based on ideas of \cite{Hog:1999} and the arithmetic circuits designed,  we develop a diagonal operator circuit which has a special structure, that is the phases are quadratic functions of the coordinates, with polynomial cost and depth. Using same techniques, other powers, instead of quadratic, can be achieved.

The gates used in our proposed designs are the ones introduced in \cite{Di:2013} where also physical implementation directions are given. The proposed designs use extensively various rotation gates. As the rotation angles of the gates vary with the size of the circuit and also small angles are required, implementation and fault tolerance issues are addressed in Appendix B, using results of the binary quantum case. 

Many of the arithmetic multilevel quantum designs of this manuscript are direct extension or modifications of binary quantum designs appeared in  \cite{Dra:1998,Bea:2003,Pav:2014} which use QFT before applying the rotation gates to one of the two integer operands and then applying the inverse QFT to bring back the result in the computational basis. The following arithmetic units are presented:

\begin{itemize}
\item Adder of two integers of $q$ qudits with depth of $O(q)$ and width of $2q$ qudits  (Subsection \ref{subsec:Adder}).

\item Adder of an integer of $q$ qudits with a constant integer. Its depth is $O(q)$ including the direct and inverse QFT blocks or $O(1)$ without the QFT blocks. Its  width is $q$ qudits  (Subsection \ref{subsec:AdderC}).

\item Single state controlled adder of an integer of $q$ with a constant. The adder is enabled if the control qudit is in a particular basis state of the different $d$ possible states, otherwise it acts as an identity. Its depth is $O(q)$ and its width is $q+1$ qudits  (Subsection \ref{subsec:CAdderC}).

\item Generalized controlled adder of an integer of $q$ qudits with a constant. It adds a multiple of the constant to the integer. The multiple depends on the state of control qudit, being between $0$ and $d-1$. Its depth is $O(q)$ and its width is $q+1$ qudits (Subsection \ref{subsec:GCAdderC}).

\item Multiplier with constant and accumulator.  It multiplies an integer of $q$ qudits with a constant and adds the product to a second integer of $q$ qudits. Its depth is $O(q)$ and its width is $2q$ qudits (Subsection \ref{subsec:MAC}).

\item Multiplier with constant. It multiplies an integer of $q$ qudits with a constant provided that the constant is relative prime with $d^q$, which is always the case when $p$ is prime.  Its depth is $O(q)$ and its width is $2q$ qudits of which $q$ qudits are ancilla initialized to the zero state and then are reset back to the zero state (Subsection \ref{subsec:MULC}).

\item Multiplier of two integers and accumulator. It multiplies two integer of $q$ qudits and adds the product to a third integer of $q$ qudits. Its depth is $O(q^2)$  and its width is $3q$ qudits (Subsection \ref{subsec:MMAC}).

\item Squarer/Multiplier with constant/Accumulator. It performs the transform $|x\rangle |z\rangle \rightarrow |x\rangle  |z + \gamma x^{2}\rangle $, where $\gamma$ is the integer constant. Its depth is $O(q^2)$ and its width is $4q$ qudits   (Subsection \ref{subsec:SMAC}).

\item General diagonal operator. It operates diagonally on a general superposition state of  $q$ qudits and changes the phases of the superposition amplitudes by applying the matrix  
$\sum_{k=0}^{d^q-1} e^{\frac{i2\pi}{d^q}f(k)} |k\rangle \langle k|$, where $f(k)$ is a function of $k$. The specific diagonal operator presented here is based on the previous squarer and some other blocks as it applies the function $f(k)=\gamma k^2$. It can be generalized for other powers of $k$ or even for polynomial functions on $k$ by exploiting similar techniques. It has a depth of $O(q^2)$ and its width is $4q$ qudits of which $3q$ qudits are ancilla (Section \ref{sec:DiagonalOperators}). 
\end{itemize}

Detailed complexity analysis in terms of quantum cost and depth is given in section \ref{sec:Complexity}, where the parameter $d$ enters the previous rough approximations. This is because many gates like the basic rotation gates used and introduced in subsection \ref{subsubsec:Rk} are synthesized using more elementary gates with a cost (and consequently depth) which depends on the dimension $d$ of the qudits.

\section{Elementary and Basic Gates on Qudits}\label{sec:Elementary}
\noindent 
We followed a hierarchical bottom-up approach to design the arithmetic circuits. At the lowest level, \emph{elementary} gates operating in a two dimensional subspace of the $d$-dimensional space of a qudit are used. Upon them, more complex gates (which are \emph{basic} for the designs) operating in the whole $d$-dimensional space are built. Some of the basic gates are reported in \cite{Di:2013}, while others like the generalized controlled and doubly controlled rotation gates are introduced here (subsection \ref{subsubsec:Rk}, subsection \ref{subsec:MMAC}  and Appendix A ).

\subsection{Generalized X gates}\label{subsec:Xjk}
\noindent
The $X^{(jk)}$ gates \cite{Di:2013} operate on a two-dimensional subspace of a $d$-level qudit by exchanging the basis states $|j\rangle,|k\rangle$,   and leaving intact the other basis states, thus they are a generalization of the well  known $X$ gate for qubits which exchanges the basis states   $|0\rangle$ and $|1\rangle$. They are defined by the $d\times d$ matrix 

\begin{equation}\label{eq:Xjk}
\begin{array}{lr}
X^{(jk)}=|j \rangle \langle k| + |k \rangle \langle j| +
\sum\limits_{\begin{subarray}{l}
n=0 \\
n \neq j \\
n \neq k
\end{subarray}}^{d-1} |n \rangle \langle n|
&
j,k=0 \ldots d-1
\end{array}
\end{equation}

It holds that $X^{(jk)}= X^{(kj)}$, so there are $d(d-1)/2$ different such gates in this family.

\subsection{Rotation gates of two levels}\label{subsec:Rxyz}
\noindent
These gates perform a rotation on a two dimensional subspace \cite{Di:2013} of a $d$-level qudit and are defined as 

\begin{equation}\label{eq:Rxyz}
\begin{array}{lrr}
R_{a}^{jk}(\theta)=exp(-i\theta \sigma _{a}^{(jk)}/2   ),
&
0 \leq j,k \leq d-1,
&
a \in \{x,y,z\}
\end{array}
\end{equation}

\noindent  where $\sigma_{x}^{(jk)}=|j\rangle \langle k| +  |k\rangle \langle j|$,
    $\sigma_{y}^{(jk)}=-i|j\rangle \langle k| +  i|k\rangle \langle j|$ and
    $\sigma_{z}^{(jk)}=|j\rangle \langle j| - |k\rangle \langle k|$ for $j,k=0\ldots d-1$ are 
matrices of dimensions $d \times d$. Parameter $\theta$ is the rotation angle, while $i=\sqrt{-1}$. 

\subsection{Generalized Controlled X gates}\label{subsec:GCX}
\noindent
The $GCX_{(m)}^{(jk)}$ gates are generalization in the qudits of the CNOT gates acting on qubits \cite{Di:2013}. Thus, they are gates which operate on a control and a target qudit. A GCX gate has three parameters, $m$, $j$ and $k$, which define its operation. A $GCX_{(m)}^{(jk)}$ acts like a
$X^{(jk)}$ on the target qudit iff the control qudit is on the basis state $|m \rangle$. Consequently, the definition matrix of such a gate is block diagonal with dimension $d^{2}\times d^{2}$ consisting of $d$  blocks of $d\times d$ dimensions each and it is given by

\begin{equation}\label{eq:GCXjk1}
\begin{array}{lr}
GCX_{(m)}^{(jk)}= |m \rangle \langle m| \otimes 
\left( |j \rangle \langle k| + |k \rangle \langle j| +
 \sum\limits_{
 \begin{subarray}{l}
 n=0 \\
 n \neq j \\
 n \neq k
 \end{subarray}}^{d-1} |n \rangle \langle n| \right) +
 \sum\limits_{
 \begin{subarray}{l}
 n=0 \\
 n \neq m 
 \end{subarray}}^{d-1} |n \rangle \langle n| \otimes I_{d}
&
j,k,m=0 \ldots d-1
\end{array}
\end{equation}

\noindent   where $I_{d}$ is the identity matrix of dimensions $d\times d$. Equation \eqref{eq:GCXjk1} can be equivalently written as

\begin{equation}\label{eq:GCXjk2}
GCX_{m}^{(jk)}= diag(I_{d},I_{d},\ldots,\underset{  m\textnormal{-th block}}{X^{(jk)}} ,\ldots,I_{d}  )
\end{equation}
\noindent   
\subsection{Hadamard gate}\label{subsec:Hadamard}

\noindent
The Hadamard gate $H^{(d)}$  on $d$-level qudits is defined by the matrix

\begin{equation}\label{eq:Hd}
\begin{split}
H^{(d)}= & \frac{1}{\sqrt{d}} 
\begin{bmatrix}
1 & 1 & 1 & 1 & \\
1 & e^{i2\pi\frac{1}{d}} & e^{i2\pi\frac{2}{d}} & \cdots & e^{i2\pi\frac{d-1}{d}} \\
1 & e^{i2\pi 2 \frac{1}{d}} & e^{i2\pi 2 \frac{2}{d}} & \cdots & e^{i2\pi 2 \frac{d-1}{d}} \\
\vdots & \vdots  & \vdots & \ddots & \vdots \\
1 & e^{i2\pi(d-1)\frac{1}{d}} & e^{i2\pi (d-1)\frac{2}{d}}&\cdots & e^{i2\pi (d-1)\frac{d-1}{d}}
\end{bmatrix} =  \\
 & \frac{1}{\sqrt{d}} 
\begin{bmatrix}
1 & 1 & 1 & 1 & \\
1 & e^{i2\pi(0.1)} & e^{i2\pi(0.2)} & \cdots & e^{i2\pi(0.d-1)} \\
1 & e^{i2\pi 2(0.1)} & e^{i2\pi 2 (0.2)} & \cdots & e^{i2\pi 2 (0.d-1)} \\
\vdots & \vdots  & \vdots & \ddots & \vdots \\
1 & e^{i2\pi(d-1)(0.1)} & e^{i2\pi (d-1)(0.2)} & \cdots & e^{i2\pi (d-1)(0.d-1)}
\end{bmatrix}
\end{split}
\end{equation}

In the above equation the notation $(0.n)$ is the fractional representation of $n/d$ in the base-$d$ arithmetic system. The application of the $H^{(d)}$ gate  to a basis state $|j \rangle$  is shown below

\begin{equation}\label{eq:Hd2basis}
\begin{split}
H^{(d)}|j \rangle   = & \frac{1}{\sqrt{d}} 
\begin{bmatrix}
1 & e^{i2\pi(0.j)} & e^{i2\pi 2(0.j)} & \ldots  & e^{i2\pi(d-1)(0.j)}  
\end{bmatrix}
^{T} = \\
 & \frac{1}{\sqrt{d}} (   |0\rangle+e^{i2\pi(0.j)}|1\rangle+\cdots +e^{i2\pi(d-1)(0.j)} |d-1\rangle   )
\end{split}
\end{equation}
\noindent   

The Hadamard gate for qudits essentially performs the order-$d$ Fourier transform, likewise the Hadamard gate for qubits performs the order-2 Fourier transform. Methods for implementation of the $H^{(d)}$  gate are proposed in \cite{Mut:2002,Erm:2007}.

\begin{figure}[!t]
\centerline{\epsfig{file=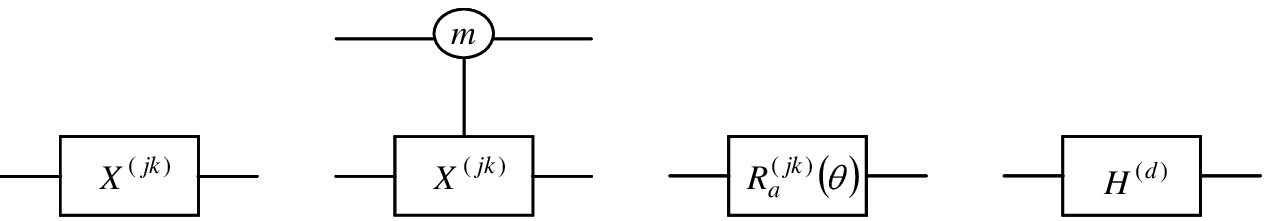, width=14cm}} 
\vspace*{13pt}
\fcaption{\label{fig:elemGates} Symbols of $X^{(jk)}$, $GCX_{(m)}^{(jk)}$, $R_{a}^{jk}(\theta)$ and $H^{(d)}$ elementary gates ($a$ is $x$,$y$ or $z$).}
\end{figure}

The symbols that will be used throughout the text for the three families of elementary gates defined and the $H^{(d)}$  gate are shown in Figure \ref{fig:elemGates}. 

\subsection{Diagonal Gates of one and two qudits}\label{subsec:D-CD}

\noindent
The qudit elementary gates of the previous section affect a 2-dimensional subspace of the whole $d$-dimensional Hilbert space of a single qudit. In this section single and two qudit diagonal basic gates affecting the whole $d$-dimensional space of one of the qudits are described and synthesized using elementary gates of the previous section.

\subsubsection{Diagonal $D^{\prime}(a_{1},a_{2},\ldots,a_{d-1})$ and $D(\varphi_{1},\varphi_{2},\ldots,\varphi_{d-1})$ gates }\label{subsubsec:D}
\noindent

The diagonal $D^{\prime}(a_{1},a_{2},\ldots,a_{d-1})$  gate \cite{Di:2013} is defined by the equation 

\begin{equation}\label{eq:Dprime}
D^{\prime}(a_{1},a_{2},\ldots,a_{d-1})=e^{i\varphi}
diag( e^{-i(a_{1}+a_{2}+\ldots +a_{d-1})} , e^{ia_{1}}, e^{ia_{2}}, \ldots , e^{ia_{d-1}} )
\end{equation}
\noindent   

It can be easily proved that such a gate can be constructed by sequentially applying $d-1$    $R_{z}^{(jk)}(\theta)$ gates as shown in the following equation 

\begin{equation}\label{eq:DprimeRz}
D^{\prime}(a_{1},a_{2},\ldots,a_{d-1})=e^{i\varphi}
R_{z}^{(01)}(a_{1}) R_{z}^{(02)}(a_{2}) \cdots  R_{z}^{(0(d-1))}(a_{d-1}) 
\end{equation}
\noindent   

A related gate is the $D(\varphi_{1},\varphi_{2},\ldots,\varphi_{d-1})$ defined as 

\begin{equation}\label{eq:D}
D(\varphi_{1},\varphi_{2},\ldots,\varphi_{d-1}) =
diag(1,e^{i\varphi_{1}},e^{i\varphi_{2}},\ldots, e^{i\varphi_{d-1}} )
\end{equation}
\noindent   

The $D(\varphi_{1},\varphi_{2},\ldots,\varphi_{d-1})$  gate is identical with the  $D^{\prime}(a_{1},a_{2},\ldots,a_{d-1})$ gate if we set 

\begin{equation}\label{eq:DDprime}
\begin{array}{cc}
a_{j}= \varphi_{j}-\frac{1}{d}\sum_{k=1}^{d-1} \varphi_{k} & j=1 \ldots d-1
\end{array}
\end{equation}

\noindent   
and add a global phase of angle $\varphi=\frac{1}{d}\sum_{k=1}^{d-1} \varphi_{k}$ to every diagonal element of 
$D^{\prime}(a_{1},a_{2},\ldots,a_{d-1})$.

\subsubsection{Controlled Diagonal $CD^{\prime}(a_{1},a_{2},\ldots,a_{d-1})$ and $CD(\varphi_{1},\varphi_{2},\ldots,\varphi_{d-1})$ gates }\label{subsubsec:CD}

\noindent
The diagonal gates of the previous subsection can be extended to operate on two qudits, where the first is the control qudit and the second is the target qudit, in the following manner: A diagonal gate  $D^{\prime}(a_{1},a_{2},\ldots,a_{d-1})$   or   $D(\varphi_{1},\varphi_{2},\ldots,\varphi_{d-1})$  is applied on the target qudit iff the control qudit is in state $|m\rangle$  , otherwise no operation is effective on the target. Thus, the $d^{2}\times d^{2}$ matrices representing such gates have the following block diagonal form 

\begin{equation}\label{eq:CDprime}
CD_{m}^{\prime}(a_{1},a_{2},\ldots,a_{d-1})=
diag(I_{d}, \ldots ,I_{d},  \underset{m\textnormal{-th} \quad \textnormal{block}}{D^{\prime}(a_{1},a_{2} ,\ldots,a_{d-1})} ,I_{d}, \ldots ,I_{d}  )
\end{equation}

\noindent   
and

\begin{equation}\label{eq:CD}
CD_{m}(\varphi_{1},\varphi_{2},\ldots,\varphi_{d-1})=
diag(I_{d}, \ldots ,I_{d},  \underset{m\textnormal{-th} \quad \textnormal{block} }{D(\varphi_{1},\varphi_{2} ,\ldots,\varphi_{d-1})} ,I_{d}, \ldots ,I_{d}  )
\end{equation}
\noindent   

\begin{figure}[h]
\centerline{\epsfig{file=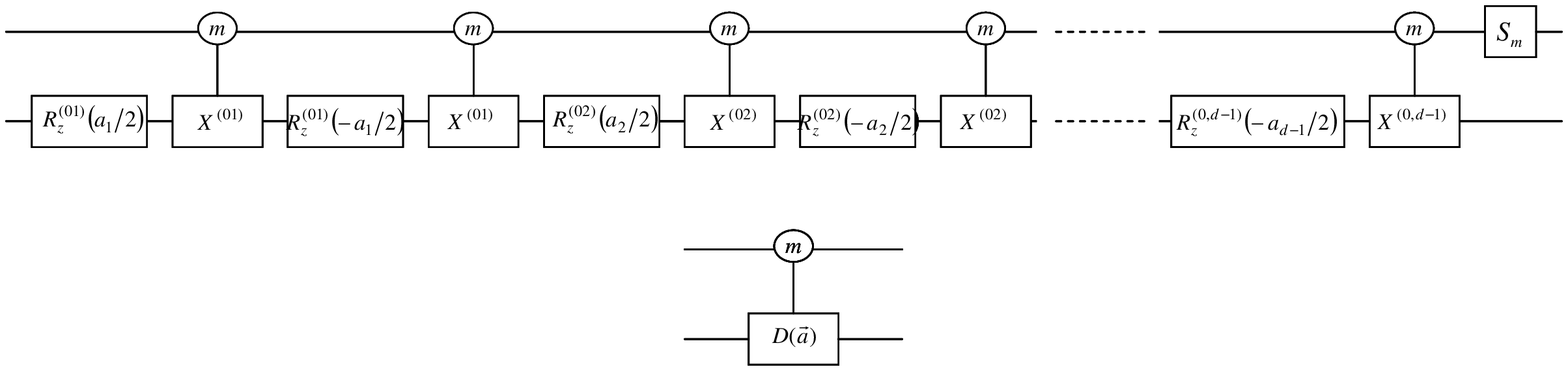, width=14cm}} 
\vspace*{13pt}
\fcaption{\label{fig:CDgate} Controlled diagonal $CD_{m}^{\prime}(a_{1},a_{2},\ldots,a_{d-1})$ gate construction and its symbol. The parameter $\overset{\rightarrow}{a}$ inside the symbol represents the angles $(a_{1},a_{2},\ldots,a_{d-1})$.  }
\end{figure}

A construction of a $CD_{m}^{\prime}(a_{1},a_{2},\ldots,a_{d-1})$ gate using $4(d-1)$ elementary $GCX_{(m)}^{(jk)}$ and  $R_{z}^{(jk)}(\theta)$ gates is shown in 
Figure \ref{fig:CDgate}. Single qudit gate  $S_{m}=diag(1, \ldots ,1, \underset{m\textnormal{-th}\,  \textnormal{pos}}{e^{i\varphi}},1,\ldots , 1)$ is a phase gate which is identical to a $D'$ gate up to a global phase.


\subsection{Generalized Controlled Rotation gate $R_{k}^{(d)}$  }\label{subsubsec:Rk}

\noindent
The controlled diagonal gates $CD_{m}^{\prime}$ and $CD_{m}$ of the previous subsection are activated whenever the control state is equal to one of the $d$ possible basis states, e.g. $|m\rangle$. We define a basic controlled diagonal gate, $R_{k}^{(d)}$, such that each one of the $d$ possible control states have a different effect on the target qudit. Such gates will be useful in the QFT and arithmetic circuits presented in the following sections.  
The $R_{k}^{(d)}$  gate is parametrized by the integer $k$. The matrix defining this gate is block diagonal of the form 

\begin{equation}\label{eq:Rk}
R_{k}^{(d)}=diag \left( 
\left( \Phi_{k}^{(d)} \right) ^{0}, 
\left( \Phi_{k}^{(d)} \right) ^{1},  \ldots ,
\left( \Phi_{k}^{(d)} \right) ^{d-1}
\right)
\end{equation}

\noindent 
where the matrix $\Phi_{k}^{(d)} $ is diagonal too, and defined with

\begin{equation}\label{eq:Phik}
\Phi_{k}^{(d)}=diag \left( 
1, 
e^{i\varphi_{1}},   
e^{i\varphi_{2}}, \ldots ,
e^{i\varphi_{(d-1)}}
\right)
\end{equation}
\noindent 

The angles $\varphi_{1},\varphi_{2},\ldots,\varphi_{(d-1)}$ depend on the parameter $k$ as follows 

\begin{equation}\label{eq:phiangles}
\begin{array}{cc}
\varphi_{m}=\frac{2\pi}{d^{k}}m, & m=1,\ldots,d-1
\end{array}
\end{equation}
\noindent

The $R_{k}^{(d)}$  gates can be equivalently written in a more detailed form consisting of a sum of tensor products of the basis states of the two qudits as 

\begin{equation}\label{eq:Rkfull}
R_{k}^{(d)}=
\sum_{j=0}^{d-1} \sum_{m=0}^{d-1} 
e^{i\frac{2\pi}{d^{k}}jm}
|j\rangle \langle j| \otimes |m\rangle \langle m| =
\sum_{j=0}^{d-1} \sum_{m=0}^{d-1} 
e^{i2\pi (0.\underbrace{00\ldots 0}_{k-1}j)m}
|j\rangle \langle j| \otimes |m\rangle \langle m|
\end{equation}
\noindent

We can see by inspecting Eq. \eqref{eq:Rkfull}  that an $R_{k}^{(d)}$  gate is a generalization on qudits of the controlled rotation gates $R_{k}=R_{z}(2\pi/2^k)=diag(1,1,1,e^{i2\pi/2^k})$   for the qubit case (where $d=2$) and this generalization will be exploited when constructing the QFT and various arithmetic circuits based on the QFT. To understand this, it is useful to see what is the effect of an $R_{k}^{(d)}$ gate when the control qudit is on a basis state $|j_{1} \rangle$ ($j_{1}=0,1,\ldots,d-1$) and the target qudit is in a superposition of equal amplitudes, but with different phases, such as  
$|b\rangle=\frac{1}{\sqrt{d}}\sum_{l=0}^{d-1}e^{i\varphi_{l}}|l\rangle$. The joint state of the two qudits after the application of the $R_{k}^{(d)}$ gate is 

\begin{equation}\label{eq:Rkeffect}
\begin{split}
R_{k}^{(d)} \left( |j_{1}\rangle |b \rangle \right)= & \frac{1}{\sqrt{d}} 
\sum_{j=0}^{d-1} \sum_{m=0}^{d-1} 
e^{i2\pi (0.\underbrace{00\ldots 0}_{k-1}j)m}
|j\rangle \underbrace{\langle j|j_{1} \rangle}_{=\delta_{jj_{1}}} 
\otimes 
 \underbrace{|m\rangle \langle m| \sum_{l=0}^{d-1}e^{i\varphi_{l}}|l\rangle}
_{=e^{i\varphi_{m}}|m\rangle}
 =  \\
 & \frac{1}{\sqrt{d}} 
\sum_{m=0}^{d-1} 
e^{i2\pi (0.\underbrace{00\ldots 0}_{k-1}j_{1})m}
|j_{1}\rangle e^{i\varphi_{m}} |m\rangle
 =  \\
 & \frac{1}{\sqrt{d}} 
 |j_{1}\rangle \sum_{m=0}^{d-1} 
e^{i2\pi \left[ (0.\underbrace{00\ldots 0}_{k-1}j_{1})m +\varphi_{m} \right] }
|m\rangle
\end{split}
\end{equation}
\noindent   

\begin{figure}[ht]
\centerline{\epsfig{file=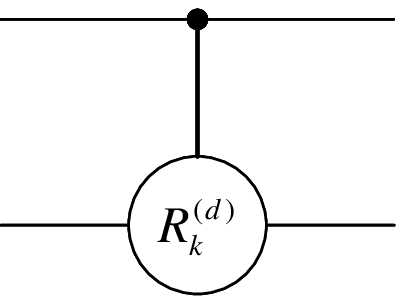, width=3cm}} 
\vspace*{13pt}
\fcaption{\label{fig:Rkdsymbol} Symbol of the generalized controlled rotation gate $R_{k}^{(d)}$.}
\end{figure}

Thus, an angle $2\pi  (0.\underbrace{00\ldots 0}_{k-1}j_{1})m=\frac{2\pi}{d^{k}} j_{1}  m$ is added to every component $|m\rangle$ of the target qudit superposition and this angle is proportional to the value $|j_{1} \rangle$ of the control qudit and also proportional to the $|m\rangle$ component of target qudit superposition.

The implementation of an $R_{k}^{(d)}$ can be achieved by sequentially combining $d-1$ controlled diagonal gates $CD_{m}(\varphi_{1},\varphi_{2},\ldots,\varphi_{d-1})$ for $m=1 \ldots d-1$ and different angles for each case of $m$ as shown below (see also Eqs. \eqref{eq:CD},\eqref{eq:Rk} and \eqref{eq:Phik} ) 

\begin{equation}\label{eq:Rkconstr}
\begin{split}
R_{k}^{(d)} = &
CD_{(1)} \left( 
\frac{2\pi}{d^{k}},\frac{2\pi}{d^{k}}2, \ldots , \frac{2\pi}{d^{k}}(d-1)
\right) \cdot
CD_{(2)} \left( 
\frac{2\pi}{d^{k}}2,\frac{2\pi}{d^{k}}4, \ldots , \frac{2\pi}{d^{k}}2(d-1)
\right) \cdots \\
 &
CD_{(d-1)} \left( 
\frac{2\pi}{d^{k}}(d-1),\frac{2\pi}{d^{k}}(d-1)2, \ldots , \frac{2\pi}{d^{k}}(d-1)(d-1)
\right)
\end{split}
\end{equation}
\noindent

Taking into account that a  $CD_{(m)}(\varphi_{1},\varphi_{2},\ldots,\varphi_{d-1})$ gate is composed by $4(d-1)$ elementary $GCX_{(m)}^{(jk)}$ and $R_{z}^{(jk)}(\theta)$ gates, then we conclude that an $R_{k}^{(d)}$ gate requires $4(d-1)^{2}$ elementary gates.  The symbol used for the $R_{k}^{(d)}$  gate in this text is shown in Figure \ref{fig:Rkdsymbol}.

\section{Quantum Fourier Transform}\label{sec:QFT}

\noindent 
The Quantum Fourier Transform on the $N$-dimensional computational basis  $\{ |0\rangle ,|0\rangle  , \ldots ,|N-1 \rangle \}$  is defined by

\begin{equation}\label{eq:QFT}
|j\rangle \xrightarrow{\text{QFT}_{N}} 
\frac{1}{\sqrt{N}}
\sum_{k=0}^{N-1} 
e^{\frac{i2\pi}{N}jk}|k\rangle
\end{equation}

Using $q$ qudits of $d$ levels \cite{Mut:2002},\cite{Erm:2007}, and setting $N=d^{q}$, the $q$ qudits basis consists of $|j\rangle = |j_{1}\ldots j_{q} \rangle = |j_{1}\rangle \ldots |j_{q}\rangle$ where for $l$-th qudit it holds $|j_{l}\rangle \in \{ |0\rangle , \ldots , |d-1 \rangle \}$. Then, the QFT action on a basis state $|j\rangle$ ($j=0 \ldots d^q-1$) is

\begin{equation}\label{eq:QFT2}
\begin{split}
\begin{split}
|j\rangle = |j_{1}\ldots j_{q} \rangle \xrightarrow{\text{QFT}_{N}}  &
\frac{1}{\sqrt{N}}
\sum_{k_{1}=0}^{d-1} 
\cdots
\sum_{k_{q}=0}^{d-1} 
e^{\frac{i2\pi}{d^{q}}j \sum_{l=1}^{q}k_{l}d^{q-l}  }
|k_{1}\ldots k_{q}\rangle 
= \\
 &
\sum_{k_{1}=0}^{d-1} 
\cdots
\sum_{k_{q}=0}^{d-1} 
\bigotimes_{l=1}^{q} 
e^{i2\pi jk_{l}d^{-l}} |k_{l}\rangle
= \\
 &
\bigotimes_{l=1}^{q} 
\sum_{k_{l}=0}^{d-1}  
e^{i2\pi jk_{l}d^{-l}} |k_{l}\rangle
= 
\end{split} \\
\left( 
\sum_{m=0}^{d-1}
e^{i2\pi (0.j_{q}) m} |m\rangle 
\right)
\left( 
\sum_{m=0}^{d-1}
e^{i2\pi (0.j_{q-1}j_{q}) m} |m\rangle 
\right)
\cdots
\left( 
\sum_{m=0}^{d-1}
e^{i2\pi (0.j_{1}j_{2}\ldots j_{q-1}j_{q}) m} |m\rangle 
\right)
\end{split}
\end{equation}
\noindent

\begin{figure}[hb]
\centerline{\epsfig{file=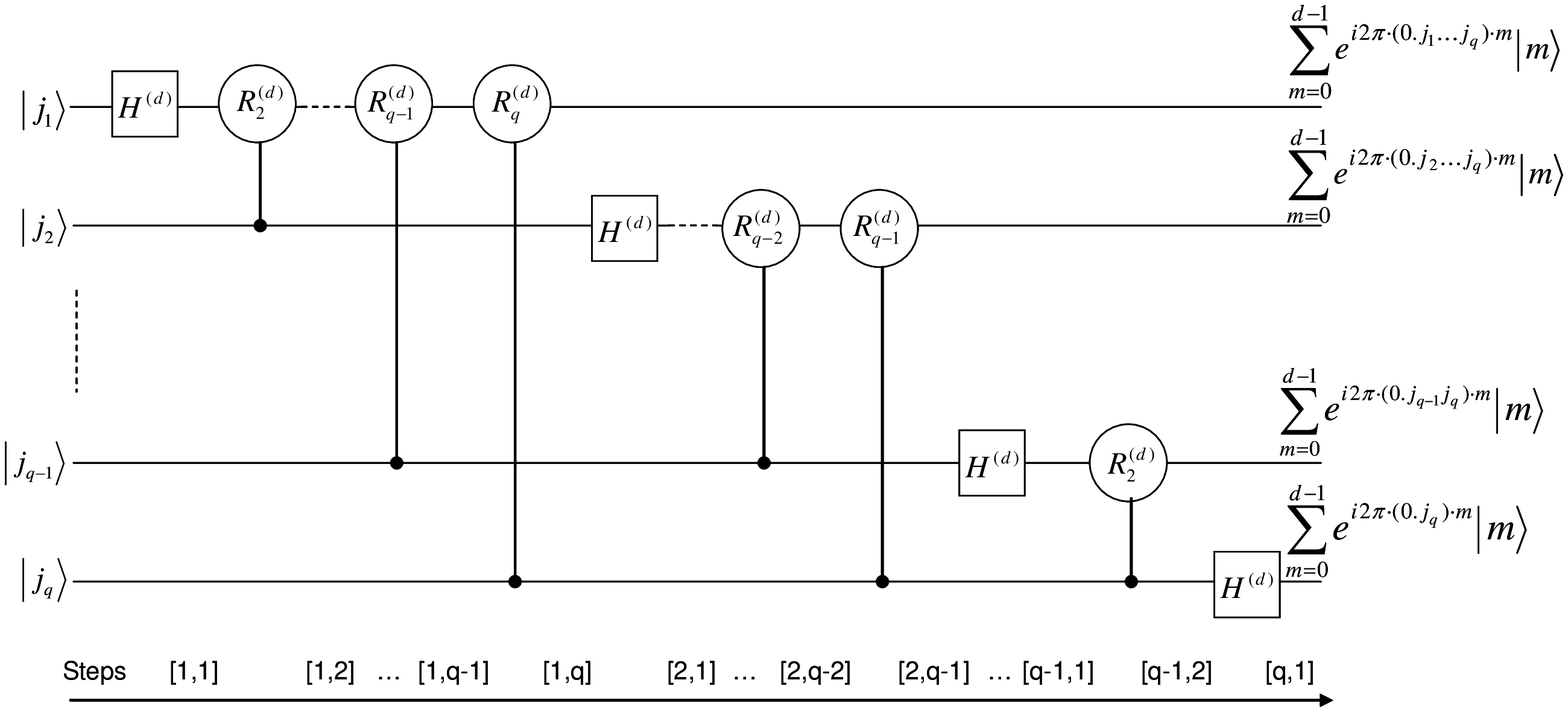, width=14cm}} 
\vspace*{13pt}
\fcaption{\label{fig:QFT} QFT circuit implemented on $d$-level qudits.}
\end{figure}

The $d$-ary representation $(j_{1}j_{2}\ldots j_{q})$ of the integer $j=j_{1}d^{q} +j_{2}d^{q-1}+ \cdots +j_{q}$ as well as the fractional $d$-ary representation $(0.j_{1}j_{2\ldots}j_{q})= j_{1}/d +j_{2}/d^{2}+ \ldots +j_{q}/d^{q}$ are used in the above definition. This tensor product form is similar to the form of the QFT of order $2^{n}$ implemented using $n$ qubits of two levels. Thus, the structure of a QFT circuit implemented with qudits is similar to the binary QFT case as depicted in Figure \ref{fig:QFT}.

Indeed, comparing the state $\sum_{m=0}^{d-1}
e^{i2\pi (0.j_{l}j_{l+1}\ldots j_{q-1}j_{q}) m} |m\rangle $ 
of the $l$-th qudit after the transformation of Eq.\eqref{eq:QFT2} with Eq.\eqref{eq:Hd2basis} and \eqref{eq:Rkeffect} we can conclude that this state can be generated by applying at the basis state $|j_{l}\rangle$  of the $l$-th qudit a Hadamard gate $H^{(d)}$  and a sequence of $q-l$ generalized rotation gates $R_{k}^{(d)}$ , with $k=2\ldots q-l+1$,  controlled by the qudits $l+1 \ldots q$, respectively. At the end, the order of the qudits must be reversed with swap gates as in the case of the QFT operated on qubits. This swapping of the qudits is not shown in Figure \ref{fig:QFT}. The inverse QFT circuit is derived by reversing horizontally  the direct QFT circuit of Figure \ref{fig:QFT} (including the SWAP gates not shown) with  opposite signs in the angles of the rotation gates.

\noindent

\section{Arithmetic Circuits}\label{sec:Arithmetic}

\noindent
The integer arithmetic circuits presented in this section are developed in a bottom up succession, starting from the simpler ones and proceeding gradually to more complex ones. The arithmetic operation are assumed to be modulo $d^q$ where $d$ are the qudit levels and $q$ is the number of qudits used to represent the integers. All the adders can be easily converted to subtractors  by using opposite sign in the angles of the rotation gates while retaining the same circuit structure.

\subsection{Adder of two integers (ADD)}\label{subsec:Adder}

\begin{figure}[!b]
\centerline{\epsfig{file=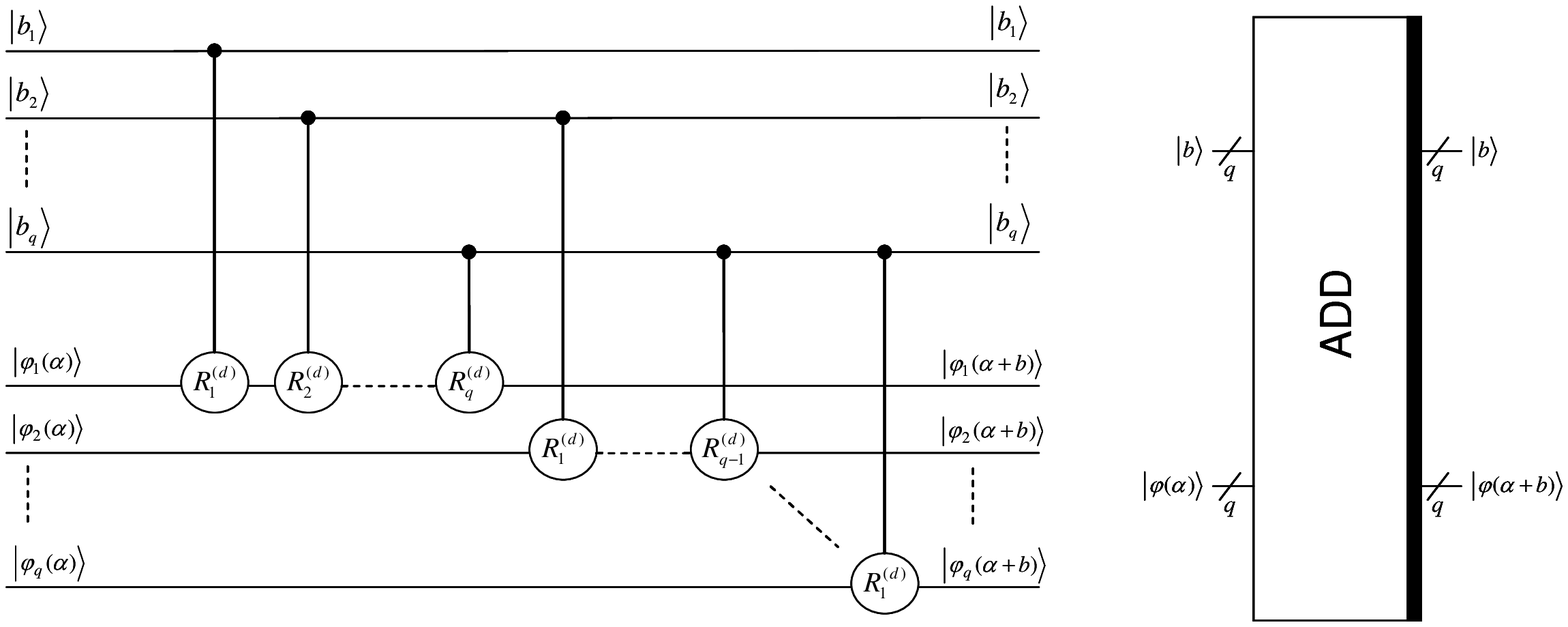, width=14cm}} 
\vspace*{13pt}
\fcaption{\label{fig:Adder} Adder of two  integers (ADD) and the respective symbol.}
\end{figure}

\noindent
A basic arithmetic operation block is an adder of two integers of $q$ $d$-ary digits each, e.g $a=(a_{1}a_{2}\ldots a_{q})$ and $b=(b_{1}b_{2}\ldots b_{q})$ or two superpositions of integers. Following  the previous sections, the most significant $d$-ary digit of an integer is indexed with $1$ while the least significant digit is indexed with $q$.
The circuit in Figure \ref{fig:Adder}  operates on $2q$ qudits, the state $|b_{1}\ldots b_{q}\rangle$  of the $q$ upper qudits (upper register) represents integer $b$ while the state of the lower $q$ qudits (lower register) represents the Fourier transformed state of the other integer $a$, that is $|\varphi_{1}(a)\rangle |\varphi_{2}(a)\rangle  \cdots |\varphi_{q}(a)\rangle$, where $|\varphi_{l}(a)\rangle= 
\sum_{m=0}^{d-1}e^{i2\pi ( 0.a_{l}a_{l+1}\ldots a_{q})m   } $ (see Eq. \eqref{eq:QFT2}). It is a generalization on qudits of the circuit proposed in \cite{Dra:1998}.
 
The first qudit of the lower register is initially in the state $|\varphi_{1}(a)\rangle $. The effect of the first rotation gate $R_{1}^{(d)}$ controlled by state $|b_{1}\rangle$ to this qudit (step[1,1]), taking into account Eq. \eqref{eq:Rkeffect}, is to evolve it  in the state 

\begin{equation}\label{eq:step11}
|\varphi_{1}(a)\rangle \xrightarrow{R_{1}^{(d)}} |\varphi_{1}(a)\rangle _{1,1} =
\frac{1}{\sqrt{d}}
\sum_{m=0}^{d-1}
e^{i2\pi \left[ \left( 0.a_{1}a_{2}\ldots a_{q}\right) +\left( 0.b_{1} \right) \right]m}   |m\rangle
\end{equation}
\noindent

The effect of the second gate $R_{2}^{(d)}$  controlled by $|b_{2}\rangle$   is to further evolve it (step[1,2]) in the state

\begin{equation}\label{eq:step12}
|\varphi_{1}(a)\rangle _{1,1} \xrightarrow{R_{2}^{(d)}} |\varphi_{1}(a)\rangle _{1,2} =
\frac{1}{\sqrt{d}}
\sum_{m=0}^{d-1}
e^{i2\pi \left[ \left( 0.a_{1}a_{2}\ldots a_{q}\right) +\left( 0.b_{1} \right) + \left( 0.0b_{2} \right) \right]m}   |m\rangle
\end{equation}
\noindent

Proceeding in a similar way up to gate $R_{q}^{(d)}$  controlled by $|b_{q}\rangle$, we find the final state (step[1,q]) of the first qudit which becomes

\begin{equation}\label{eq:step1q}
|\varphi_{1}(a)\rangle _{1,q-1} \xrightarrow{R_{q}^{(d)}} |\varphi_{1}(a)\rangle _{1,q} =
\frac{1}{\sqrt{d}}
\sum_{m=0}^{d-1}
e^{i2\pi \left[ \left( 0.a_{1}a_{2}\ldots a_{q}\right) +\left( 0.b_{1}b_{2} \ldots b_{q} \right)   \right]m}   |m\rangle
\end{equation}
\noindent

In general, the final state of the $l$-th qudit of the lower register is found to be

\begin{equation}\label{eq:stepl}
|\varphi_{l}(a)\rangle _{l,q-l+1} =
\frac{1}{\sqrt{d}}
\sum_{m=0}^{d-1}
e^{i2\pi \left[ \left( 0.a_{l}a_{l+1}\ldots a_{q}\right) +\left( 0.b_{l}b_{l+1} \ldots b_{q} \right)   \right]m}   |m\rangle
\end{equation}
\noindent

Applying Eq. \eqref{eq:stepl} to each lower register qudits we can find that the  lower register has the final joint state 

\begin{equation}\label{eq:stepend}
|\varphi(a)\rangle _{1} |\varphi(a)\rangle _{2} \cdots |\varphi(a)\rangle _{q} \xrightarrow{ADD}
\bigotimes_{l=1}^{q} 
\frac{1}{\sqrt{d}}
\sum_{d=0}^{d-1}
e^{i2\pi \left[ \left( 0.a_{l}a_{l+1}\ldots a_{q}\right) +\left( 0.b_{l}b_{l+1} \ldots b_{q} \right) +  \right]m}   |m\rangle =
|\varphi (a+b) \rangle
\end{equation}
\noindent

This is the quantum Fourier transform of the sum state $|a+b  \pmod{d^{q}}\rangle$. By applying the inverse QFT at the lower register we can get the desired sum in the computational basis, while the upper register remains in the initial state $|b \rangle$. The required direct and inverse QFT blocks are not shown in Figure \ref{fig:Adder}. 

\subsection{Adder of an integer with constant (ADDC$_{b}$)}\label{subsec:AdderC}

\noindent
Whenever one of the integers is constant, e.g. $b=(b_{1}b_{2}\ldots b_{q})$, then the upper register in Figure \ref{fig:Adder} is not necessary and all the controlled rotation gates become single qudit rotation gates  with their angles defined by the constant integer $b$. Thus (see Eqs. \eqref{eq:Rk} and \eqref{eq:Phik}), we must apply on the $l$-th qudit of the lower register  a sequence of $q-l+1$ rotation gates $ \left( \Phi_{k}^{(d)} \right) ^{b_{k+l-1}} = \sum_{m=0}^{d-1} e^{i\frac{2\pi}{d^{k}}mb_{k+l-1}} |m\rangle \langle m| $, for $k=1\ldots q-l+1$. This product of gates can be merged into one gate of the form

\begin{equation}\label{eq:addcBgate}
\begin{split}
B_{l}(b) = &
\prod_{k=1}^{q-l+1} \left( \Phi_{k}^{(d)} \right) ^{b_{k+l-1}} 
=\\
&
\prod_{k=1}^{q-l+1} \left(  
\sum_{m=0}^{d-1} e^{\frac{i2\pi m}{d^{k}}} |m\rangle \langle m|
\right) ^{b_{k+l-1}}
=\\
&
\sum_{m=0}^{d-1} \left(  
\prod_{k=1}^{q-l+1}
e^{\frac{i2\pi m}{d^{k}}b_{k+l-1}}
\right) |m\rangle \langle m|
\end{split}
\end{equation}
\noindent

\begin{figure}[!t]
\centerline{\epsfig{file=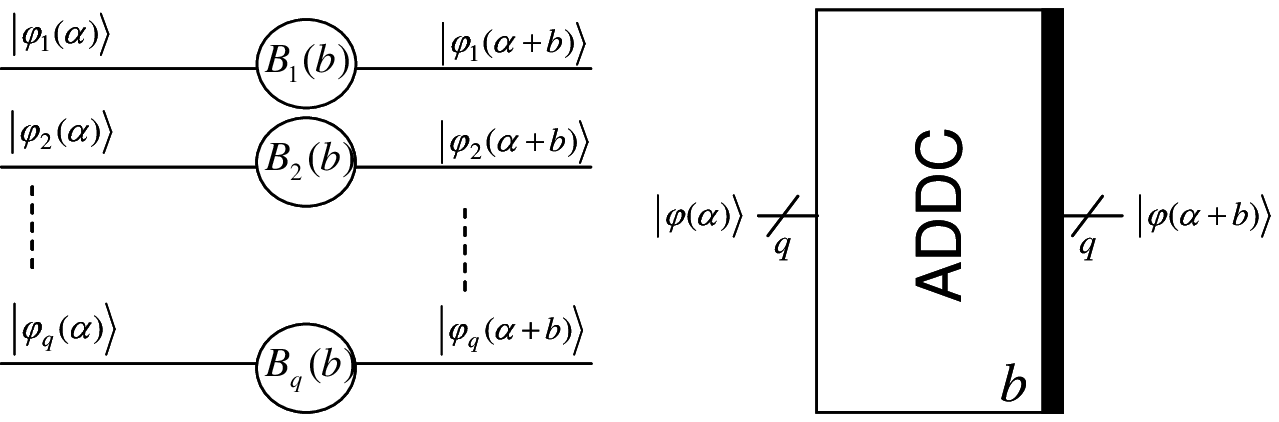, width=10cm}} 
\vspace*{13pt}
\fcaption{\label{fig:AdderC} Adder of an integer with constant $b$ (ADDC$_{b}$) and the respective symbol.}
\end{figure}

These are diagonal gates of the form of Eq. \eqref{eq:D}, and their angles depend on the constant $b$ by the relation 

\begin{equation}\label{eq:addcBangle}
\varphi_{l,m}(b)=
\sum_{k=1}^{q-l+1}
\frac{2\pi}{d^{k}}mb_{k+l-1}
\end{equation}

\noindent
so they can be constructed with elementary $R_{z}^{(jk)}(\theta)$  gates  using the procedure described in  subsection \ref{subsec:D-CD}. Figure \ref{fig:AdderC} shows the constant $b$ adder (direct and inverse QFT blocks not included in the diagram). Likewise the general adder ADD, this adder performs the addition modulo $d^{q}$.

\subsection{Single State Controlled Adder of an integer with constant (C$_{c}$ADDC$_{b}$) }\label{subsec:CAdderC}

\begin{figure}[hb]
\centerline{\epsfig{file=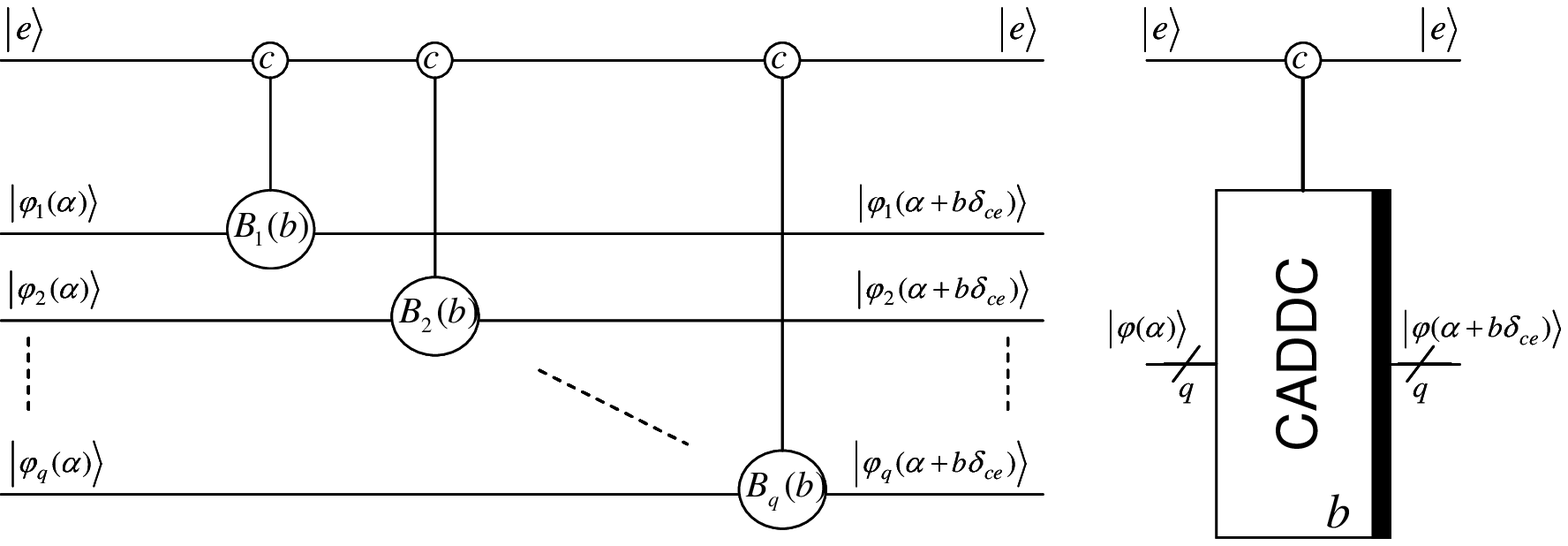, width=14cm}} 
\vspace*{13pt}
\fcaption{\label{fig:CAdderC} State $|c\rangle$  controlled adder with constant $ b$ (C$_{c}$ADDC$_{b}$) and the respective symbol.}
\end{figure}

\noindent
The constant adder ADDC$_{b}$ can be easily converted to a constant adder controlled by the state  of an additional control qudit so as to perform the transformation

\begin{equation}\label{eq:CADDC}
C_{c}ADDC_{b}\left( |e\rangle |a\rangle \right)=
|e\rangle |a+b\delta_{ce}\rangle
\end{equation}
\noindent

where $\delta_{ce}$ is the Kronecker delta function. Consequently, the addition is performed iff the control state  equals $|c\rangle$, otherwise the target state   $|a\rangle$ remains unaltered. The one state controlled constant adder C$_{c}$ADDC$_{b}$ can be constructed as shown in Figure \ref{fig:CAdderC} if the one qudit rotation gates $B_{l}(b)$ of Figure \ref{fig:AdderC} are converted to the respective two qudits diagonal gates controlled by state $|c\rangle$. These gates are exactly the $CD_{(c)}$ gates of  subsection \ref{subsec:D-CD}.

\noindent

\subsection{Generalized Controlled Adder of an integer with constant (GCADDC$_{b}$) }\label{subsec:GCAdderC}

\noindent
A useful generalization of the previous C$_{c}$ADDC$_{b}$ circuit can be achieved if we permit all the basis states of the control qudit to have an influence on the result of the addition. Such a circuit will be named Generalized Controlled Adder with constant $b$ and is defined by  the relation

\begin{equation}\label{eq:GCADDC}
GCADDC_{b}\left( |e\rangle |a\rangle \right)=
|e\rangle |a+be\rangle
\end{equation}
\noindent

The above equation can be rewritten as 

\begin{equation}\label{eq:GCADDC2}
GCADDC_{b}\left( |e\rangle |a\rangle \right)=
|e\rangle |a+b\delta _{1e}+2b\delta _{2e}+ \cdots  +(d-1)b\delta _{(d-1)e} \rangle
\end{equation}
\noindent

\begin{figure}[h]
\centerline{\epsfig{file=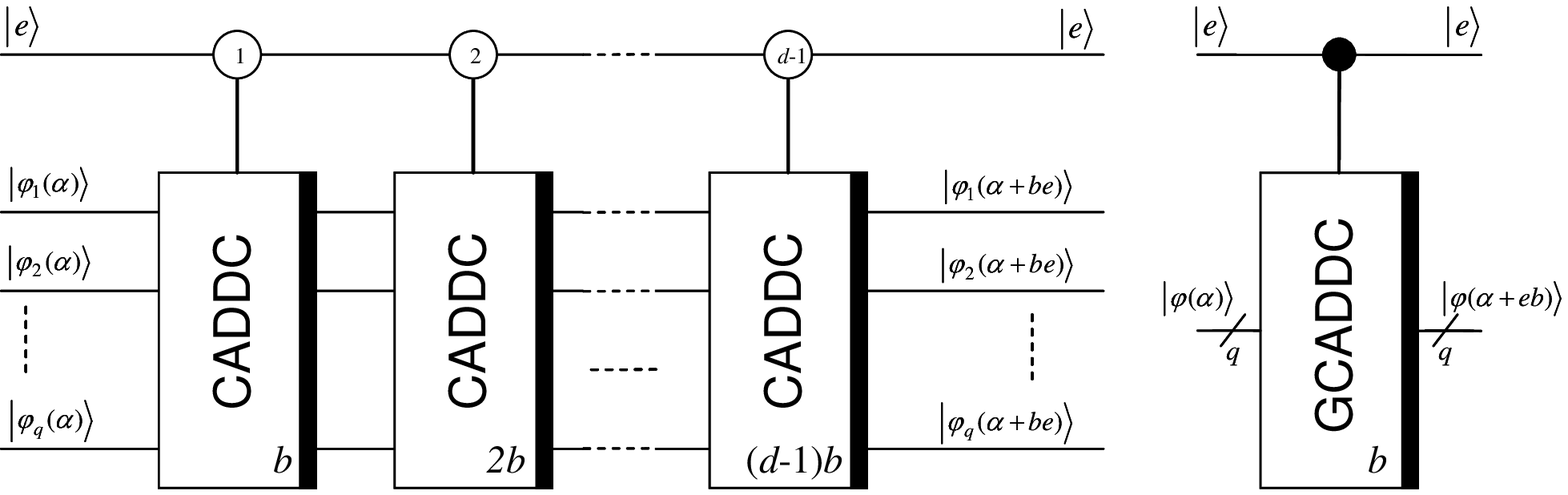, width=14cm}} 
\vspace*{13pt}
\fcaption{\label{fig:GCAdderC} Generalized controlled adder with constant $b$ (GCADDC$_{b}$) and the respective symbol.}
\end{figure}

Equation \eqref{eq:GCADDC2} directly leads to the implementation of Figure \ref{fig:GCAdderC} where $d-1$ consecutive applications of C$_{c}$ADDC$_{bc}$ ($c=1\ldots d-1$) adders are employed. 

\noindent

\subsection{Multiplier with constant and Accumulator (MAC$_{b}$) }\label{subsec:MAC}

\noindent
A Multiplier with constant and Accumulator MAC$_{b}$ multiplies a $q$ qudits integer $x$ with a constant $b$ of $q$ $d$-ary digits, and accumulates the product $bx$ to a $q$ qudits integer $a$ (modulo $d^{q}$). Namely, the MAC$_{b}$ circuit consists of two $q$ qudits registers holding initially the states $|x\rangle$  and $|a\rangle$  and transforms them as 

\begin{equation}\label{eq:MAC}
MAC_{b}\left( |x\rangle |a\rangle \right)=
|x\rangle |a+bx\rangle
\end{equation}
\noindent

Taking into account that $x$ can be written as $(x_{1}x_{2} \ldots x_{q})=\sum_{l=1}^{q}x_{l}d^{q-l}$ then Eq. \eqref{eq:MAC} can be written as 

\begin{equation}\label{eq:MAC2}
\begin{split}
MAC_{b}\left( |x\rangle |a\rangle \right)= &
|x\rangle |a+b\sum_{l=1}^{q}x_{l}d^{q-l}    \rangle = \\
 & |x\rangle |a + x_{q}b + x_{q-1}db + \cdots + x_{1}(d^{q-1}b) \rangle
\end{split}
\end{equation}
\noindent

\begin{figure}[!t]
\centerline{\epsfig{file=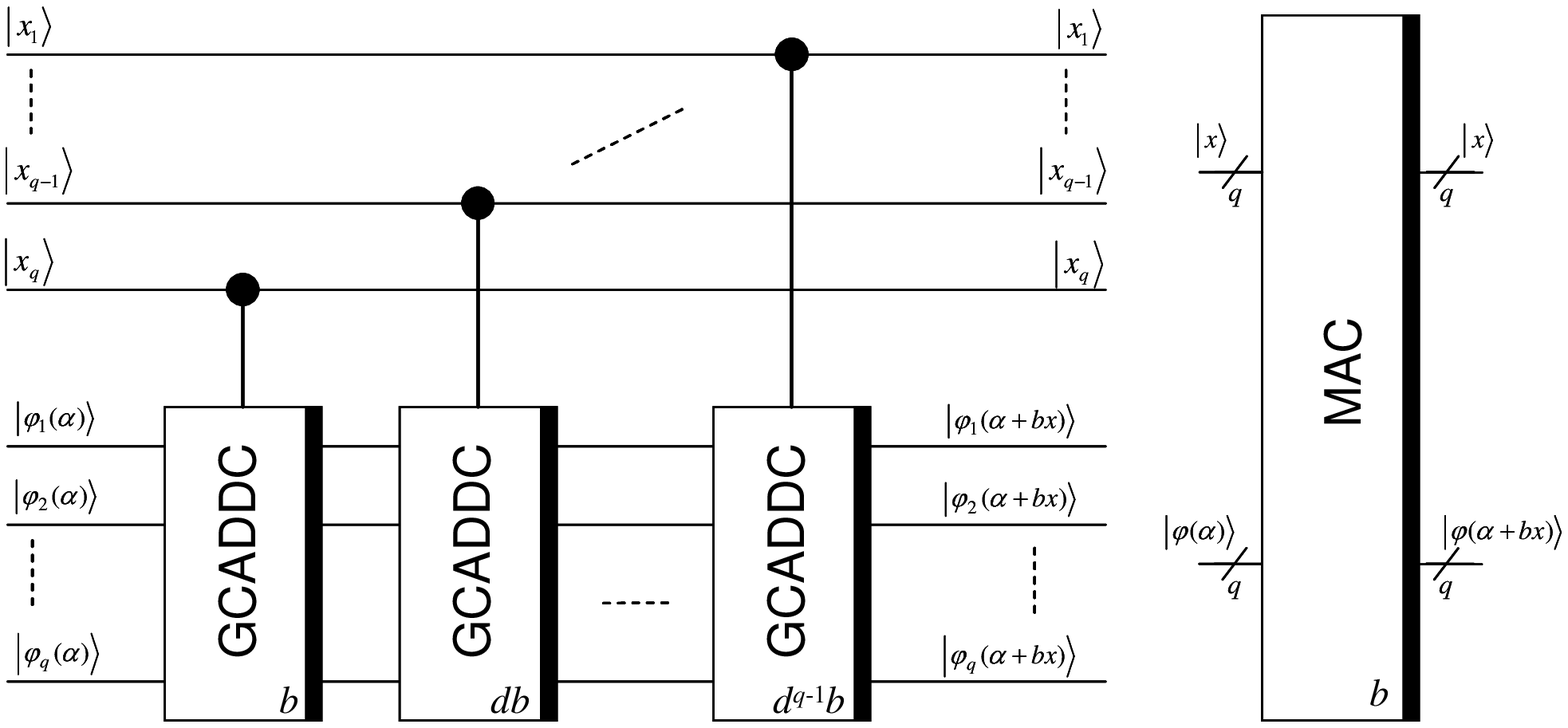, width=14cm}} 
\vspace*{13pt}
\fcaption{\label{fig:MAC} Multiplier with constant Accumulator $b$ (MAC$_{b}$) and the respective symbol.}
\end{figure}

This means that the above transformation can be implemented by applying $q$ GCADDC circuits, where  the control is done consecutively by the qudits $x_{q},x_{q-1},\ldots,x_{1}$ and the constant parameter for each one GCADDC block is $b,db,\ldots,d^{q-1}b$ (modulo $d^{q}$), respectively, as shown in Figure \ref{fig:MAC}.

\noindent

\subsection{Multiplier with constant  (MULC$_{b}$) }\label{subsec:MULC}

\begin{figure}[!b]
\centerline{\epsfig{file=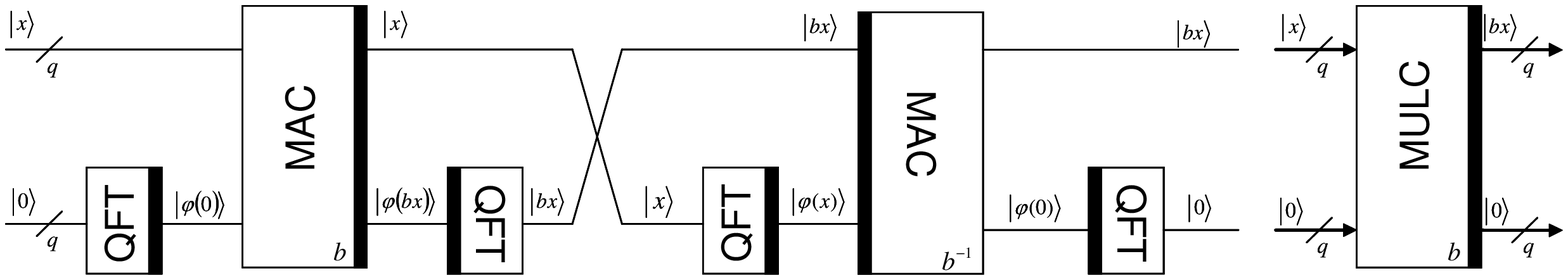, width=14cm}} 
\vspace*{13pt}
\fcaption{\label{fig:MULC} Multiplier with constant $b$ (MULC$_{b}$) and the respective symbol.}
\end{figure}

\noindent
A multiplier (modulo $d^{q}$) with constant $b$ implements the function $f:{0\ldots d^{q}-1} \rightarrow {0\ldots d^{q}-1}$ with $y=f(x)=bx \pmod{q^{q}}$. When cosntant $b$ is relative prime to $d^{q}$ then there exists the inverse $b^{-1} \pmod{d^{q}}$ and consequently there exists the inverse function $f^{-1}(y)=b^{-1}y \pmod{d^{q}} = b^{-1}bx \pmod{d^{q}}=x$. This always happens when $d$ is a prime number. 
Figure \ref{fig:MULC} shows how to construct a Multiplier with constant $b$ using two $MAC_{b}$ blocks and the necessary direct and inverse QFT blocks. It requires a $q$ qudits register initially holding the integer $x$ and another $q$ qubits ancilla register initially in zero state. At the end, one register is set to the state $|bx \pmod{d^{q}}\rangle$ while the other register is set to state zero, so effectively the ancilla register is reset back and can be reused.

\noindent

In the diagram of Figure \ref{fig:MULC}, the boxes with the black strip at their right side are the "direct" blocks while these with the black strip at their left side are the respective inverses. The operation of the inverse MAC with parameter $b^{-1}$ is to perform substraction instead of accumulation, that is referring to Figure \ref{fig:MULC},  we have the operation $MAC_{b^{-1}}^{-1} | bx\rangle |\varphi(x) \rangle = | bx\rangle |\varphi (x-b^{-1}(bx)) \rangle = | bx\rangle |0\rangle$. 
The inverse MAC$^{-1}$ has the same internal topology as the direct MAC of Figure \ref{fig:MAC} (of course with parameter $b^{-1}$ instead of $b$) with the only difference that the angles of its rotation gates have a minus sign. By inspecting the labels at the qudit buses of Figure \ref{fig:MULC} describing the respective states we can conclude that the circuit implements the multiplication 

\begin{equation}\label{eq:MULC}
MULC_{b}\left( |x\rangle |0\rangle \right)= |bx\rangle |0\rangle
\end{equation}
\noindent

Excluding the ancilla register, which is in the zero state before and after the operation and thus it remains unentangled, we conclude that this circuit performs the desired multiplication operation.

\section{Diagonal Operators on $q$ qudits}\label{sec:DiagonalOperators}

\noindent
The diagonal operator on $q$ qudits of $d$ levels, as its name implies, is a circuit whose unitary matrix of  dimensions $d^{q}\times d^{q}$
has a diagonal form. The circuit developed in this section is such that the diagonal elements of the matrix are integer powers of the principal root of unity $e^{\frac{i2\pi}{d^q}}$ and the integer powers are a function $f(k)$ of the coordinate $k=1\ldots d^{q}-1$ of the elements. In what follows, the diagonal operator circuit developed is for the function $f(k)=\gamma k^{2}$, where $\gamma$ is an integer constant. Thus, the definition of our diagonal operator on $q$ qudits is  

\begin{equation}\label{eq:diagOp}
\Delta_{\gamma}^{(q)}=\sum_{k=0}^{Q-1} e^{\frac{i2\pi}{Q}f(k)} |k\rangle \langle k|
\end{equation}

\noindent
where $Q=d^{q}$ and $f(k)=\gamma k^{2} \pmod{d^{q}}$. All the diagonal entries of the above matrix are integer powers of the basic phase $\omega =e^\frac{i2\pi}{Q} $. The effect of this matrix upon a general superposition state of $q$ qubits will be 

\begin{equation}\label{eq:diagOpaEff}
\sum_{k=0}^{Q-1} c_{k} |k \rangle \xrightarrow{\Delta_{\gamma}^{(q)}} 
\sum_{k=0}^{Q-1} c_{k} e^{\frac{i2\pi}{Q}f(k)}  |k \rangle
\end{equation}
\noindent

The circuit that implements the operator of Eq. \eqref{eq:diagOp} will be derived by exploiting results of \cite{Hog:1999} which are given for the case of binary quantum circuits. A prerequisite for this construction is a Squarer/Multiplier with constant/Accumulator circuit (SMAC) that computes the function $f$  involving two $q$ qudits registers as in 

\begin{equation}\label{eq:SMAC}
SMAC_{\gamma}(|k\rangle |z\rangle ) = |k\rangle |z + f(k) \rangle  =
|k\rangle |z + \gamma k^{2} \pmod{Q} \rangle 
\end{equation}
\noindent

Such an SMAC circuit will be described in  subsection \ref{subsec:SMAC}. 

\begin{figure}[h]
\centerline{\epsfig{file=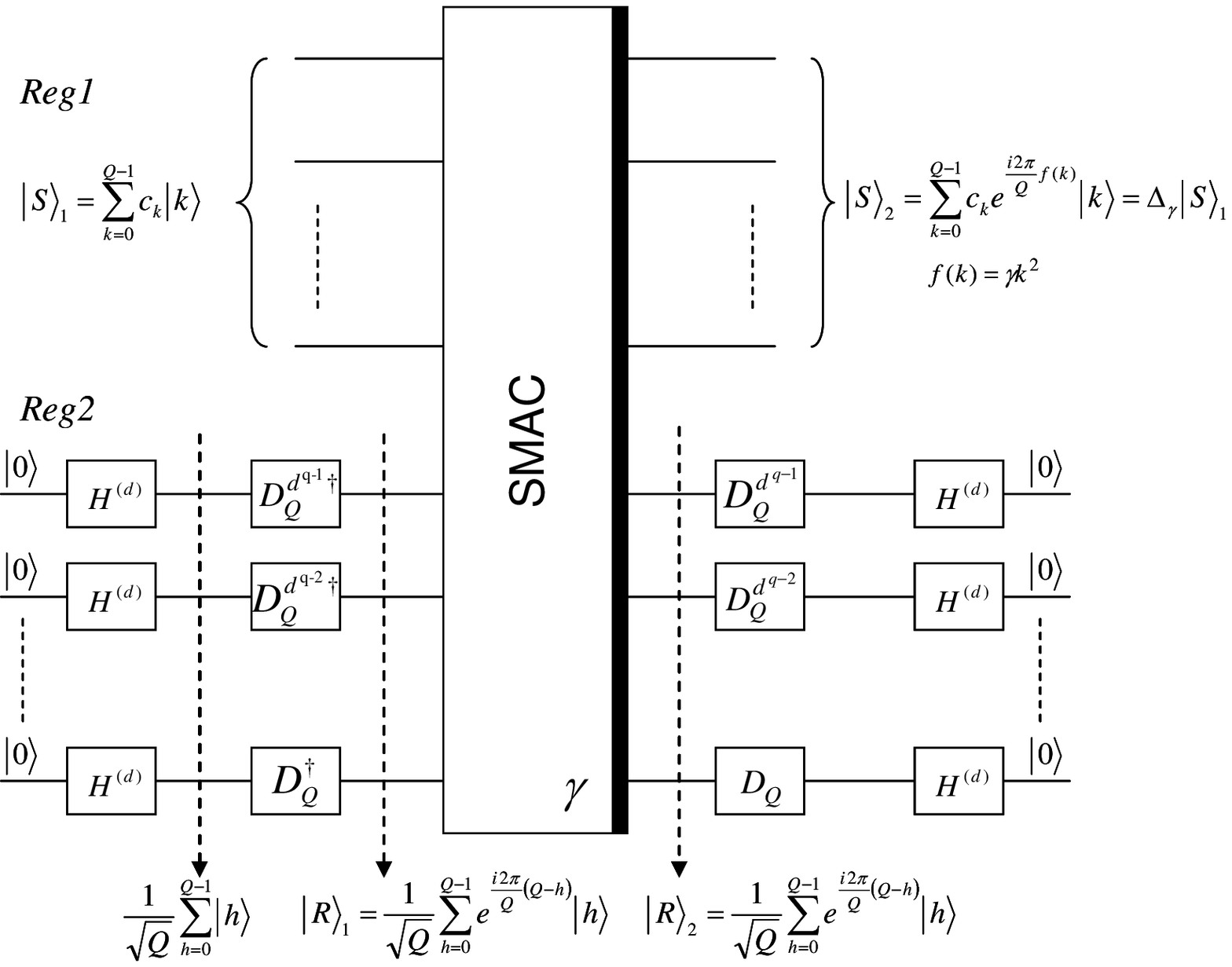, width=14cm}} 
\vspace*{13pt}
\fcaption{\label{fig:DiagOper} Diagonal $q$ qubits operator $\Delta_{\gamma}^{(q)}$.}
\end{figure}

Figure \ref{fig:DiagOper} shows the diagonal operator circuit with entries dependable on the function $f(k)=\gamma k^{2} \pmod{Q}$. Two  quantum registers are used, each $q$ qudits wide, namely \emph{Reg1} and \emph{Reg2}. The upper register \emph{Reg1} is assumed to be in a general superposition state prior the operator $\Delta_{\gamma}^{(q)}$ is applied as described in Eq. \eqref{eq:diagOpaEff}, while the lower register \emph{Reg2} is an ancilla register with zero initial and final state.

The first step is to form in the ancilla register \emph{Reg2} the uniform superposition state $|R\rangle = \frac{1}{\sqrt{Q}} \sum_{h=0}^{Q-1} |h\rangle$. This is accomplished with the application of $q$ Hadamard gates $H^{(d)}$ on each qudit of the register. 
Then, we apply on each qudit the diagonal gates $D_{Q}^{d^{m}\dagger}$ for $m=0\ldots q-1$ . The matrix representing these gates is 

\noindent

\begin{equation}\label{eq:Dqudit}
D_{Q}^{d^{m}\dagger}=
diag(1,\omega _{Q} ^{-1\cdot d^{m}},\omega _{Q} ^{-2\cdot d^{m}},\ldots ,\omega _{Q} ^{(d-1)\cdot d^{m}} ), \qquad m=0\ldots q-1
\end{equation}

\noindent
and it has exactly the same form of the diagonal gates of Eq. \eqref{eq:D}. The joint affect of these gates at \emph{Reg2} is given by their tensor product which is a diagonal matrix too, of dimensions $d^{q}\times d^{q}$ 

\begin{equation}\label{eq:Dwhole}
D_{QQ}^{\dagger}=
diag(1,\omega _{Q}  ^{-1},\omega _{Q} ^{-2}, \ldots ,\omega _{Q} ^{Q-2}, \omega _{Q} ^{Q-1} )
\end{equation}
\noindent

Then, the state of \emph{Reg2}  becomes

\begin{equation}\label{eq:R1}
|R\rangle _{1}= \frac{1}{\sqrt{Q}} = \sum_{h=0}^{Q-1} \omega ^{(Q-h)} |h \rangle
\end{equation}
\noindent

The initial state of \emph{Reg1} is assumed to be a general superposition of basis states and can be expressed as

\begin{equation}\label{eq:S1}
|S \rangle _{1}= \sum_{k=0}^{Q-1} c_{k} |k \rangle = 
\sum_{n=0}^{Q-1} \sum_{k \in K_{n}} c_{k} |k \rangle , 
\qquad K_{n}=\{k: f(k)=n   \}
\end{equation}
\noindent

In the right hand side of Eq. \eqref{eq:S1} we have grouped all the basis states with value $k$ such that $f(k)=n$ in a set $K_{n}$ and then sum over all the states belonging to sets $K_{n}$. The expediency of this grouping will be clear later.
Combining Eq. \eqref{eq:S1} and \eqref{eq:R1} we find the joint state of \emph{Reg1} and \emph{Reg2} just before the application of the SMAC block, which is given by the tensor product 

\begin{equation}\label{eq:SR1}
|S \rangle _{1} \otimes |R \rangle _{1} = 
\frac{1}{\sqrt{Q}}
 \sum_{n=0}^{Q-1} \sum_{k \in K_{n}} \sum_{h=0}^{Q-1}
c_{k} \omega ^{Q-h} |k \rangle |h \rangle
\end{equation}
\noindent

Taking into account the effect of the SMAC block given by Eq. \eqref{eq:SMAC} we get the state of the two registers after the application of the SMAC 

\begin{equation}\label{eq:SR2}
\begin{split}
|SR \rangle _{2}  = & SMAC(|S \rangle _{1} \otimes |R \rangle _{1}) = \\
 & 
\frac{1}{\sqrt{Q}}
\sum_{n=0}^{Q-1} \sum_{k \in K_{n}} \sum_{h=0}^{Q-1}
c_{k} \omega ^{Q-h} |k \rangle |h + f(k) \pmod{Q }\rangle \\
& 
\frac{1}{\sqrt{Q}}
\sum_{n=0}^{Q-1} \sum_{k \in K_{n}} \sum_{h=0}^{Q-1}
c_{k} \omega ^{Q-h} |k \rangle |h + n \pmod{Q }\rangle
\end{split}
\end{equation}
\noindent

We are going to use $m=h+n \pmod{Q}$ as the index of the inner summation in place of $h$. We observe that for a particular $n$, as $h$ takes the values from $0$ to $Q-1$, then $m=h+n \pmod{Q}$ takes one value a time ("1-1" mapping), that is the new index $m$ will be in the same range from $0$ to $Q-1$. Thus the lower and upper limits of the new index $m$ remain the same and we have $h=m-n \pmod{Q}$ and $Q-h=n+(Q-m) \pmod{Q}$. Also, it holds $\omega ^{Q}=1$. Then Eq. \eqref{eq:SR2} becomes 

\begin{equation}\label{eq:SR2b}
\begin{split}
|SR \rangle _{2}  = & 
\frac{1}{\sqrt{Q}}
\sum_{m=0}^{Q-1} \sum_{n=0}^{Q-1} \sum_{k \in K_{n}} 
c_{k} \omega ^{n} \omega ^{Q-m} |k \rangle |m\rangle \\
& 
\frac{1}{\sqrt{Q}}
\sum_{n=0}^{Q-1} \sum_{k \in K_{n}} c_{k} \omega ^{n} |k \rangle \otimes
\sum_{m=0}^{Q-1} \omega ^{Q-m}   |m \rangle  \\
&
\sum_{k=0}^{Q-1} c_{k} \omega ^{f(k)} |k \rangle \otimes
\frac{1}{\sqrt{Q}}
\sum_{m=0}^{Q-1} \omega ^{Q-m}   |m \rangle  \\
&
\left( \Delta_{\gamma}^{(q)} |S \rangle \right) \otimes |R \rangle
\end{split}
\end{equation}
\noindent

This shows that \emph{Reg1} has the desired state of Eq. \eqref{eq:diagOpaEff} and it is disentangled with respect to \emph{Reg2} which remains in state of Eq. \eqref{eq:R1}.  Thus, the ancilla \emph{Reg2} can be reset without any effect on the \emph{Reg1}. The resetting can be accomplished as shown in Figure \ref{fig:DiagOper} by applying in the reverse sequence (a) the inverse of the gates $H^{(d)}$ and (b) the inverse of $D_{Q}^{d^{m}\dagger}$,  which are the $H^{(d)\dagger}=H^{(d)*}$ (conjugate Hadamard) and $D_{Q}^{d^{m}}$, respectively. An alternative method would be to measure \emph{Reg2} and depending on the measurement result to apply GCX gates controlled by the measurement classical result. This measurement would not affect \emph{Reg1} as it is disentangled with respect to \emph{Reg2}.

\subsection{Multiplier of two integers / Accumulator (MMAC)}\label{subsec:MMAC}

\noindent 
The construction of the SMAC block requires a  multiplier of two integers and accumulator block (MMAC) whose operation is to multiply  integer $x$ with integer $y$ and accumulate the product $xy$ to  integer $z$ (modulo $d^{q}$). This means that the MMAC block is applied on three $q$ qudits registers and performs the transformation

\begin{equation}\label{eq:MMAC}
MMAC \left( |x\rangle |y\rangle |z\rangle \right) = 
|x\rangle |y\rangle |z+xy\rangle 
\end{equation}

\noindent

If $x=(x_{1}x_{2}\ldots x_{q})=\sum_{t=1}^{q} x_{t}d^{q-t}$ and $y=(y_{1}y_{2}\ldots y_{q})=\sum_{s=1}^{q} y_{s}d^{q-s}$ are the $d$-base representations of the two integers , then their product (modulo $d^{q}$) is given by 

\begin{equation}\label{eq:xy}
xy=\sum_{s=0}^{q-1} d^{s} \sum_{t=0}^{q-1}x_{q-t}y_{q-s+t}
\end{equation}
\noindent

In Eq. \eqref{eq:xy} the full product terms corresponding to powers $d^{s}$ with $s \geq q$ have not been included, because the product is to be calculated modulo $d^{q}$. Also, digits with negative index (e.g. $x_{-1}$), as well as with index greater than $q$ (e.g. $x_{q+1}$), are assumed zero.
The calculation of the product and the accumulation can be performed in an similar way as in the MAC circuit given in subsection \ref{subsec:MAC}. We assume that the state corresponding to the accumulation register integer $|z\rangle $ is already Fourier transformed and taking into account Eq. \eqref{eq:QFT2} which expresses the QFT we expect that the $l$-th qudit of the accumulation result $|z+xy\rangle $ prior the inverse QFT is 

\begin{equation}\label{eq:xyQFT}
|\varphi_{l}(z+xy)\rangle = 
|0\rangle +
e^{\frac{i2\pi}{d^{l}} (z+xy)}   |1\rangle + 
e^{\frac{i2\pi}{d^{l}} (z+xy)2}   |2\rangle + \cdots +
e^{\frac{i2\pi}{d^{l}} (z+xy)(d-1)}   |d-1\rangle
\end{equation}
\noindent

Thus, to bring and initial state $|\varphi_{l}(z)\rangle$  of the $l$-th qudit to the state of Eq. \eqref{eq:xyQFT}   we must  add various integer multiples of the basic angle $(2\pi/d^{l})$. Namely, taking into account Eq. \eqref{eq:xy}, the angles that must be added to the amplitude phases of a basis state $|r\rangle$ ($r=0\ldots d-1$) in the superposition  $|\varphi _{l}(z)\rangle$ of Eq. \eqref{eq:xyQFT} are

\begin{equation}\label{eq:xyAngles}
\Phi_{l,r} = \frac{2\pi}{d^{l}}xyr =
2\pi r  \sum\limits_{s=0}^{q-1}
d^{s-l}
\sum_{t=0}^{s}x_{q-t}y_{q-s+t}=
2\pi r \sum_{s=0}^{l-1} d^{s-l}
\sum_{t=0}^{s} x_{q-t}y_{q-s+t}
\end{equation}
\noindent

The restriction $s<l$ at the upper limit of the first sum of Eq. \eqref{eq:xyAngles} comes due to the periodicity $exp(\varphi +2\pi d^{n} )= exp(\varphi )$ that holds for any integer $d$ and any non negative integer $n$. The restriction $t\leq s$ at the upper limit of the second sum results because $y_{q-s+t}=0$ for $t>s$. Replacing with $k=l-s$, Eq. \eqref{eq:xyAngles} becomes 

\begin{equation}\label{eq:xyAngles2}
\Phi_{l,r} = 
2\pi r \sum_{k=1}^{l} d^{-k}
\sum_{t=0}^{l-k} x_{q-t}y_{q+k-l+t}
\end{equation}
\noindent

Consequently, the angles that must be added to the phase amplitude of the $|r\rangle$ component of the superposition are $(2\pi/d^{k})x_{m}y_{n}r$ and depend on indices $m=q-t$ and $n=q+k-l+t$. This can be attained if we introduce the notion of a double controlled generalized rotation gate applied to three qudits, two controls and one target.
Similarly to Eq. \eqref{eq:Rkfull} which is the definition of the simply controlled generalized rotation gate, we define the double controlled generalized rotation gate $\overline{R} _{k}^{(d)}$ with the $d^{3}\times d^{3}$ matrix

\begin{equation}\label{eq:RRk}
\overline{R} _{k}^{(d)} =
\sum_{m=0}^{d-1 } \sum_{n=0}^{d-1 } \sum_{r=0}^{d-1 }
e^{\frac{i2\pi}{d^{k}}mnr}
\left( |m\rangle \langle m| \right) \otimes 
\left( |n\rangle \langle n| \right) \otimes 
\left( |r\rangle \langle r| \right) 
\end{equation}

Figure \ref{fig:RRkdsymbol} depicts the symbol for this double controlled rotation gate. In Appendix A a construction of $\overline{R} _{k}^{(d)}$   will be presented using some of the elementary and basic gates introduced in Section 3.

\begin{figure}[!b]
\centerline{\epsfig{file=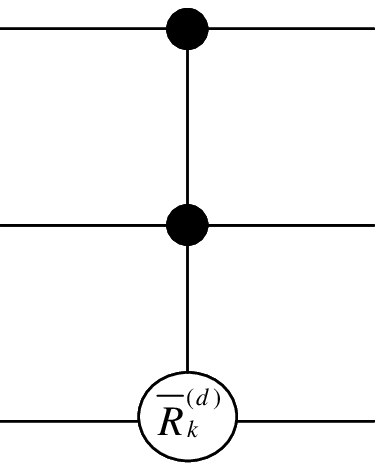, width=3cm}} 
\vspace*{13pt}
\fcaption{\label{fig:RRkdsymbol} Symbol of  $\overline{R} _{k}^{(d)}$ gate.}
\end{figure}

\begin{figure}[h]
\centerline{\epsfig{file=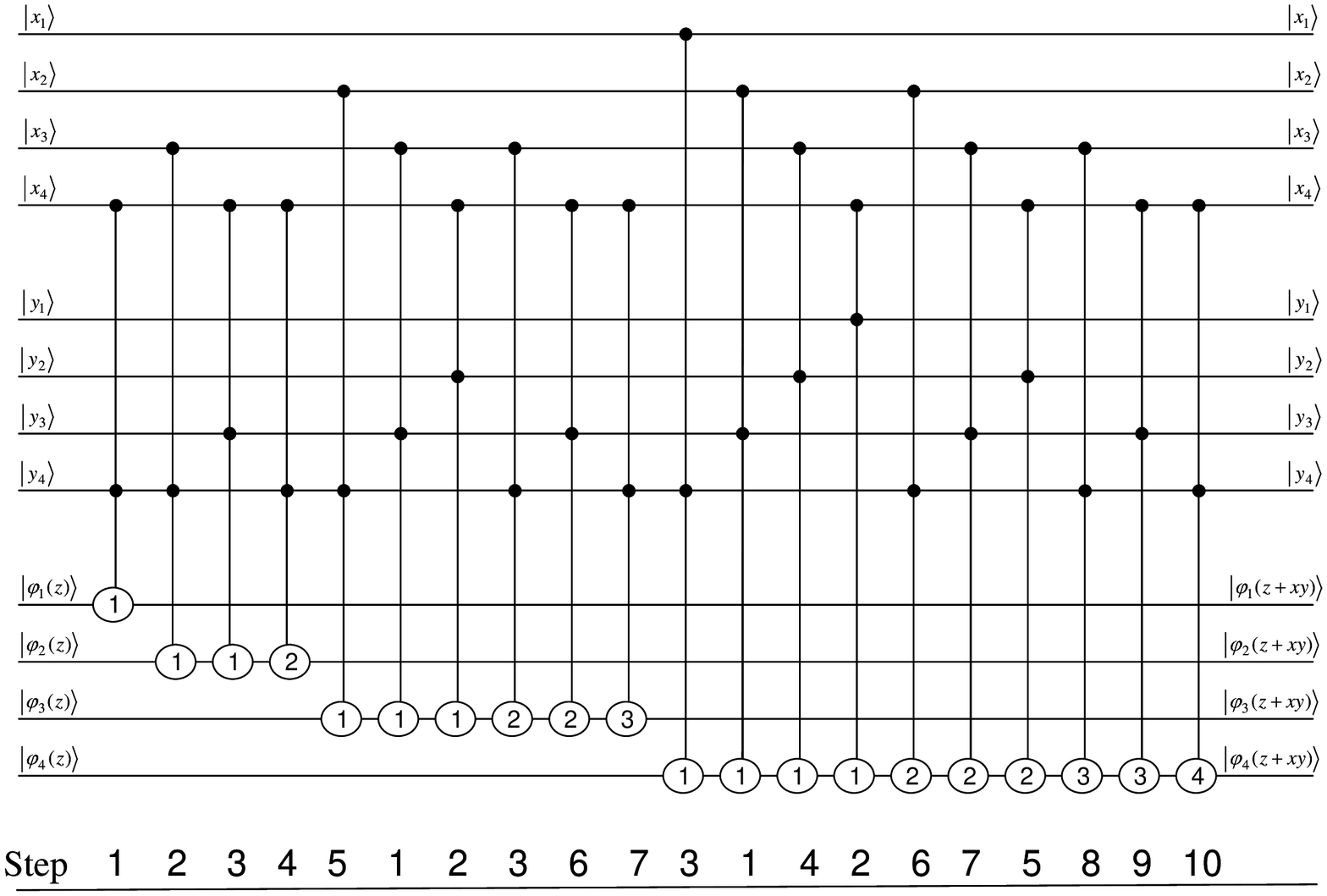, width=14cm}} 
\vspace*{13pt}
\fcaption{\label{fig:MMAC} Two 4-qudits integers Multiplier/Accumulator (MMAC) and parallel operation of gates.}
\end{figure}

The topology of the MMAC circuit can be directly concluded from Eq.\eqref{eq:xyAngles2}, as this equation describes  which  gates have to applied and which are their control connections to the qudits carrying $|x\rangle$ and $|y\rangle$. Figure \ref{fig:MMAC}  shows an example MMAC for the case of $q=4$. In this figure the $\overline{R} _{k}^{(d)}$  gates are represented with the value $k$ inside the circle. Generalization for any value of $q$ is obvious. 

We observe in Figure \ref{fig:MMAC} and in Eq. \eqref{eq:xyAngles2} that $l-k+1$ $\overline{R} _{k}^{(d)}$ gates are sequentially applied on the $l$-th target qudit for a specific $k$ ($l=1 \ldots q$,\quad $k=1 \ldots l$). In total, $C_{l}=\sum_{k=1}^{l} (l-k+1) =\frac{l(l+1)}{2}$ $\overline{R} _{k}^{(d)}$  gates are applied on the $l$-th target qudit. Summing over all target qudits we find the total number of gates used  $C_{MMAC}(q)=\sum_{l=1}^{q} C_{l} = \sum_{l=1}^{q} \frac{l(l+1)}{2} =\frac{1}{6}q^{3}+\frac{1}{2}q^{2} + \frac{1}{3}q$. The same value gives the depth of the circuit as arranged in Figure \ref{fig:MMAC}. Indeed, for the example $q=4$ we find $C_{MMAC}(4)=20$.  
We can exploit the fact that gates $\overline{R} _{k}^{(d)}$  mutually commute as they are diagonals and rearrange them so as to achieve a parallelization in their execution. The gates that can be executed simultaneously are those that operate on different qudits. An example of the proposed parallelization for the case $q=4$ is shown Figure \ref{fig:MMAC}, where below each gate is shown the soonest timestep in which it can be executed. E.g. at the first timestep three gates can be executed in parallel as none of these gates operate on the same qudit as the other two. We can generalize this parallelization scheme and conclude that we can achieve a depth of about $q(q+1)/2$ which is quadratic instead of cubic without the proposed rearrangement.

\subsection{Squarer/Multiplier/Accumulator  (SMAC)}\label{subsec:SMAC}

\begin{figure}[h]
\centerline{\epsfig{file=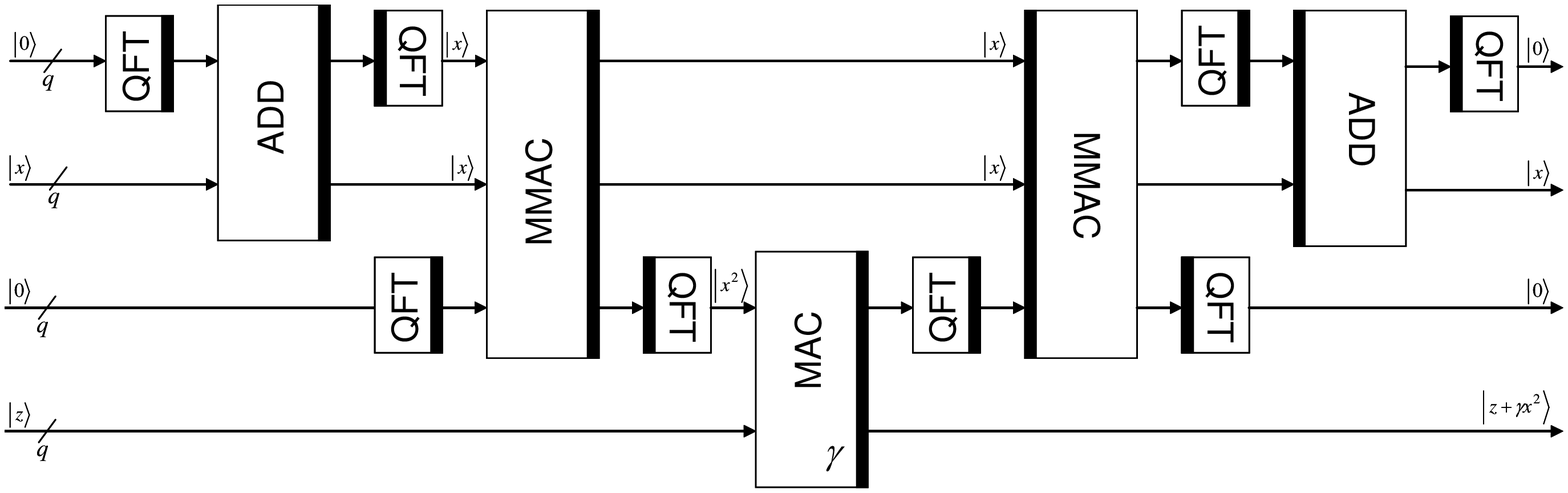, width=14cm}} 
\vspace*{13pt}
\fcaption{\label{fig:SMAC} Squarer/Multiplier with constant $\gamma$/Accumulator (SMAC$_{\gamma}$). }
\end{figure}

\noindent 
The MMAC circuit allows the  construction of the SMAC$_{\gamma}$ circuit described by Eq. \eqref{eq:SMAC} and required for the $q$ qudits diagonal operator $\Delta_{\gamma}^{(q)}$. The Squarer/Multiplier with constant $\gamma$ /Accumulator modulo $d^{q}$ is presented in block diagram in Figure \ref{fig:SMAC}. It uses $4q$ qudits, $2q$ of which are ancilla qudits with zero initial and final state, where $q$ is the number of qudits used to represent the argument $x$. The $4q$ qudits are grouped into four registers of $q$ qudits each. The second register from top holds the argument $|x\rangle$ while the bottom register holds the accumulation value $|z+\gamma x^{2} \bmod d^{q}\rangle$. 

The first step is to set the state of the top register into the same state as the second one, which is $|x\rangle$. This is accomplished with the adder block sandwiched between two QFT blocks, direct and inverse (This operation could be also achieved using a sequence of GCX gates to "copy" the second's register state to the first). In a second step, the two states $|x\rangle$ and $|x\rangle$  of the two top registers are multiplied together by the MMAC and  the product is accumulated to the third register from top, which was initially in the zero state. At this stage, the joint state of the three top registers is $|x\rangle |x\rangle |x^{2} \pmod{d^{q}} \rangle$. Next, the MAC$_{\gamma}$ block follows to multiply the constant $\gamma$ with the $|x^{2} \pmod{d^{q}} \rangle$ state of the third register. The result is accumulated to the bottom register, which was initially in state $|z\rangle$. At this point the joint state of the four registers can be described by $|x\rangle |x\rangle |x^{2} \pmod{d^{q}} \rangle |z+\gamma x^{2} \pmod{d^{q}} \rangle$.
What remains is to reset the first and the third ancilla registers. The inverse MMAC resets the third register by performing substraction instead of accumulation of the product $|x^{2}\rangle$. The inverse MMAC is constructed like the direct MMAC with opposite angles in its rotation gates. Last, the inverse adder resets the top ancilla register. Consequently, the circuit of Figure \ref{fig:SMAC} implements the transformation 

\begin{equation}\label{eq:SMAC2}
SMAC \left( |0\rangle |x\rangle |0\rangle |z\rangle \right) =
|0\rangle |x\rangle |0\rangle |z + \gamma x^{2} \pmod{d^{q}} \rangle 
\end{equation}
which is exactly the transformation of Eq. \eqref{eq:SMAC} if the ancilla registers are ignored.

\section{Complexity Analysis }\label{sec:Complexity}

\begin{figure}[!b]
\centerline{\epsfig{file=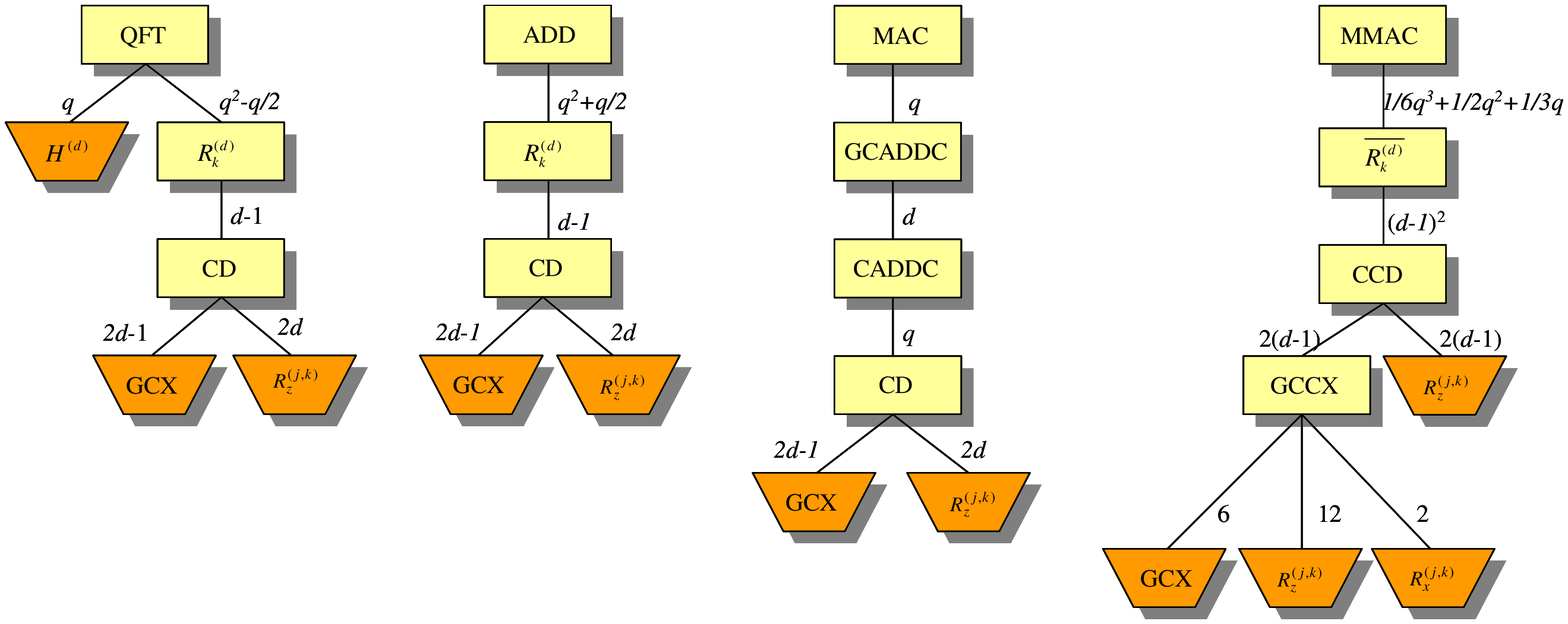, width=14cm}} 
\vspace*{13pt}
\fcaption{\label{fig:Hier}Hierarchy break-down of various arithmetic quantum circuit proposed. Elementary gates are shown in trapezoid shape as the leafs of the tree. Parameter $q$ is the arguments size, while $d$ is the dimensionality of the qudits. }
\end{figure}

\noindent
The arithmetic quantum circuits proposed in the previous sections are broken down to the level of elementary gates $H^{(d)}$, $R_{z}^{(jk)}(\theta)$, $R_{x}^{(jk)}(\theta)$ and $GCX_{m}^{(jk)}$ introduced in Section \ref{sec:Elementary}. This decomposition is depicted in Figure \ref{fig:Hier} in a tree structure, where the root of of each tree is some of the complete circuits proposed and the leaves of the tree (trapezoids) represent the elementary gates. The edges of each tree are labeled with the number of components needed by each level from one level below (no label stands for 1). The SMAC and the Diagonal operator are not included in this Figure, but their costs and depths can be easily calculated after the calculations of the blocks shown in this figure.

A rough complexity analysis in terms of quantum cost (number of elementary gates used) and depth (execution time) can be done with the help of Figure \ref{fig:Hier}. The analysis assumes that single and two qudits gates are equivalent in terms of costs and execution time. Exact costs and depths depend on the particular implementations. The total gates count for each block can be found by traversing the tree emerging from the inspected block down to each leaf of the subtree. The labels of the edges for each path are multiplied and then the products of each path used are summed together. E.g. the QFT circuit needs $q$ Hadamard gates, $(q^{2}-q/2)(d-1)(2d-1)$ $GCX_{m}^{(jk)}$gates and $(q^{2}-q/2)(d-1)2d$ $R_{z}^{(jk)}(\theta)$  gates. Similar calculations provide us with the quantum costs shown in Table \ref{table:Complexity}, which shows only the highest order terms. 

\begin{table}[!t]
\tcaption{Quantum cost, depth and width of the proposed arithmetic circuits.}
\centerline{ 
\begin{tabular}{l c c c}\\
\hline
Circuit & Cost  & Depth & Width \\\hline
QFT 	& $4d^{2}q^{2}$		& $8d^{2}q$				& $q$  \\
ADD 	& $4d^{2}q^{2}$ 	& $4d^{2}q$ 			& $2q$ \\
MAC 	& $4d^{2}q^{2}$ 	& $4d^{2}q$		& $2q$ \\
MULC 	& $24d^{2}q^{2}$ 	& $32d^{2}q$ 	& $2q$ \\
MMAC 	& $7d^{3}q^{3}$		& $21d^{3}q^{2}$	& $3q$ \\
SMAC 	& $14d^{3}q^{3}$ 	& $42d^{3}q^{2}$ 	& $4q$ \\
$\Delta _{\gamma}^{q}$ 		&$14d^{3}q^{3}$ 	& $42d^{3}q^{2}$ 	& $4q$ \\
\hline
\end{tabular}  } 
\label{table:Complexity}
\end{table}

The depth calculation will be done in more detail by finding first the depths of QFT, ADD, MAC and MMAC blocks and then the depths of MULC and SMAC. 

\begin{itemize}
\item[QFT] At first glance Figure \ref{fig:QFT} exhibits a quadratic depth $O(q^{2})$, but it can be easily shown that we can parallelize the execution with an appropriate reordering of the gates and thus achieve a linear depth, namely \emph{depth}(QFT)=$8d^{2}q$.

\item[ADD] Similarly as in the QFT case, a reordering of gates in Figure \ref{fig:Adder} offers a linear depth too, that is \emph{depth}(ADD)=$4d^{2}q$.

\item[MAC] Concurrent execution of gates is possible in this case, too. It can be easily seen that by flattening the hierarchy MAC-GCADDC-CADDC, $q$ different controlled gates $B_{l}(b)$ (Eq. \eqref{eq:addcBgate}) belonging in different GCADDC blocks can be executed concurrently. Thus, the depth of the MAC is of the order $O(4d^{2}q)$ instead of $O(4d^{2}q^{2})$ as directly calculated by the number of elementary gates.

\item[MULC] Observing Figure \ref{fig:MULC} we find \emph{depth}(MULC)=3\emph{depth}(QFT)+2\emph{depth}(MAC), as the two middle QFT blocks (direct and inverse) can be executed simultaneously. Thus, we derive \emph{depth}(MULC)= $32d^{2}q$.

\item[MMAC] The reordering of gates achieves $q(q+1)/2$ execution steps of double controlled rotation gates. Taking into account the decomposition of these three qudit gates into single and two qudits gates (see Appendix A) we end up in \emph{depth}(MMAC)=$21d^{3}q^{2}$. 

\item[SMAC] From the previous calculations and Figure \ref{fig:SMAC} we find that the dominant depth of SMAC in leading order is twice the depth of the MMAC block.

\item[$\Delta _{\gamma}^{q}$] The depth of the diagonal circuit is essentially the depth of the SMAC.

\end{itemize}

\section{Conclusions and Future Work}\label{sec:Conclusion}
\noindent
In this paper we presented an assortment of quantum circuits for multilevel qudits. These are basic integer arithmetic operations circuits (like addition, multiplication/accumulation and multiplication) as well as more complex circuits such as squarers. Additional extensions can be applied. E.g., the ADD, ADDC, MAC and MULC circuits can be converted to single qudit controlled versions. Such controlled versions could be useful to the multilevel qudits quantum phase estimation algorithms and quantum simulations. 

The general diagonal operator has been developed for the special case of a quadratic function $f(x)=\gamma k^{2}$, where $k$ is the coordinate of the diagonal element, however using the same techniques we can easily construct diagonal operators for any power of $k$ and even for a polynomial function on $k$. E.g. the Squarer/Multiplier/Accumulator can be converted to a circuit that accumulates  the third power by inserting additional MMAC units in Figure \ref{fig:SMAC}.

The designs  are based on the alternative representation of an integer after QFT transformed instead of the usual computational basis representation, a method which has been already exploited in the binary qubits case. QFT based arithmetic circuits design is a versatile method to develop many arithmetic circuits. E.g. there is no need to handle carries which leads to space reduction. Moreover, if it is suitably used, it can offer advantages in terms of speed. This is possible when similar blocks are iterated to act on a datapath whose state follows the QFT representation. The extensive usage of rotation gates (which mutually commute) on such a datapath permits their rearrangement so as they execute concurrently. This capability is observed in the MAC block, where the application of a suitable reordering of gates led to depth reduction from $O(q^{2})$ to $O(q)$. Similarly, the depth of the MMAC block reduced from $O(q^{3})$ to $O(q^{2})$.

Another advantage that has been observed in designs adopting the QFT method is their robustness to various kinds of deviations from the ideal operation. E.g. approximate QFT \cite{Coppersmith:1994} or QFT banding is the design procedure of eliminating small angle rotation gates. Studies of the Shor's algorithm which uses the QFT showed that the algorithm still works sufficiently even when a large proportion of the QFT rotation gates are eliminated \cite{Bar:1996,Fowler:2004,Nam:2012}. Recent studies extended to circuits beyond QFT. In \cite{Nam:2013a,Nam:2013b} the simultaneous gate pruning of rotation gates of the QFT circuit and the QFT based modular exponentiator of Beauregard's circuit \cite{Bea:2003} were simulated. The simulation results showed similar robustness of Shor's algorithm to these gates eliminations. This robustness is sustained even if the parameters of the remaining rotation gates are randomly selected \cite{Nam:2015}. The above results suggest that a similar robustness is expected in the multidimensional qudits case and further investigation to be carried.

On the other side, there is a drawback related to the requirement of reliable implementing high accuracy small angles rotation gates. Moreover, these gates must belong to a set of fault tolerant gates if large scale quantum computation is considered. Fortunately, as shown in Appendix B, approximation of these gates is possible, albeit with a cost. However, the remarks of the previous paragraph suggest that this cost may be much lower if approximate computation is adopted.

For the above reasons and also because the exact cost depends on the exact technology used, which for qudits is at an early stage, the complexity analysis of section \ref{sec:Complexity} is to be considered as a crude indicator of performance. Despite that, we think that the proposed designs enrich the toolkit of the future quantum computing.

\nonumsection{References}

\appendix

\noindent 
The double controlled gates $\overline{R} _{k}^{(d)}$ which are used in the MMAC block can be decomposed to single and two qudits elementary gates. Equation \eqref{eq:RRk} states that the effect of the $\overline{R} _{k}^{(d)}$  to the target qudit state is  multiplication by a diagonal matrix $d \times d$ of the form $diag(1,e^{\frac{i2\pi}{d^{k}}mn},e^{\frac{i2\pi}{d^{k}}2mn}, \ldots ,    e^{\frac{i2\pi}{d^{k}}(d-1)mn})$ iff the two control states are  $|m\rangle$ and $|n\rangle$. Consequently, to implement this gate we need (as was the case of the  gate $CD_{m}(\varphi _{1}, \ldots , \varphi _{d-1} )$), the construction of double controlled diagonal gates of the form 

\begin{equation}\label{eq:CCD}
CCD_{(m,n)}( \varphi_{1},\varphi_{2},\ldots,\varphi_{d-1})=
diag(I_{d}, \ldots ,I_{d},  \underset{mn-th \quad block}{D(\varphi_{1},\varphi_{2} ,\ldots,\varphi_{d-1})} ,I_{d}, \ldots ,I_{d}  )
\end{equation}

\noindent   
where the diagonal $D(\varphi_{1},\varphi_{2} ,\ldots,\varphi_{d-1})$ is applied to the target qudit iff the two control qudits states are $|m\rangle$ and $|n\rangle$. The angles are $\varphi _{l} =(2\pi / d^{k}mnl)$ for $l=1 \ldots d-1$. Thus, the gate $\overline{R} _{k}^{(d)}$   is constructed by successively using $(d-1)\times (d-1)$ three qudit gates $CCD_{(m,n)}(\varphi_{1},\varphi_{2},\ldots,\varphi_{d-1})$ as follows 

\begin{equation}\label{eq:RRkConstr}
\overline{R} _{k}^{(d)} =
\prod_{m=0}^{d-1} \prod_{n=0}^{d-1} CCD_{(m,n)} (1,\frac{i2\pi}{d^{k}}mn,\frac{i2\pi}{d^{k}}2mn, \ldots ,    \frac{i2\pi}{d^{k}}(d-1)mn )
\end{equation}
\noindent   

\begin{figure}[!b]
\centerline{\epsfig{file=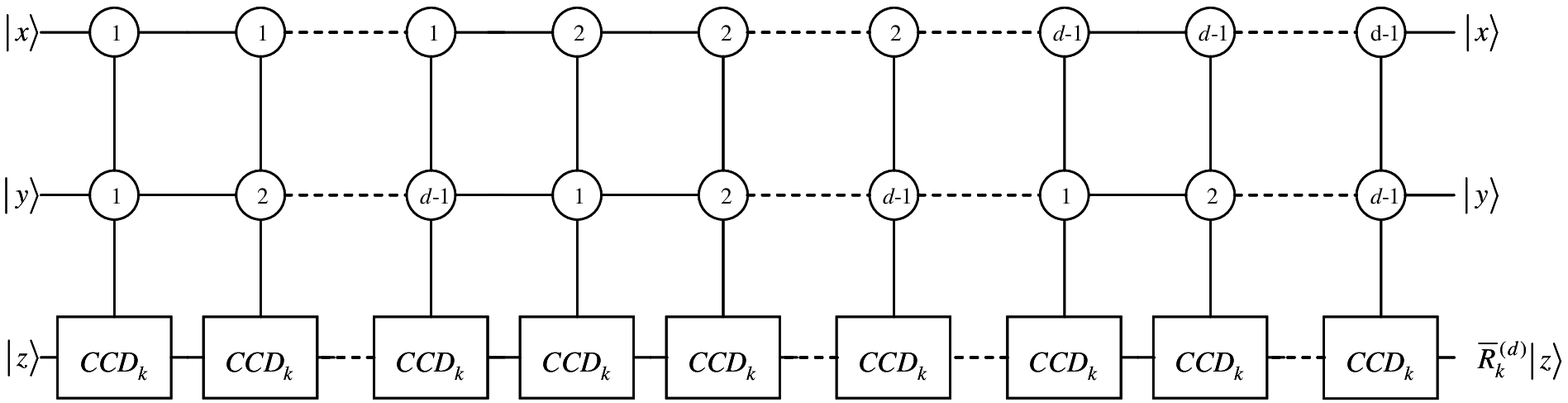, width=14cm}} 
\vspace*{13pt}
\fcaption{\label{fig:RRkDecomp}Decomposition of  $\overline{R} _{k}^{(d)}$ gate into CCD gates.}
\end{figure}

The above decomposition is depicted in Figure \ref{fig:RRkDecomp}. The parameter $k$ inside the rectangular symbol of the CCD gates corresponds to the parameter $k$ of $\overline{R} _{k}^{(d)}$ gate, while the values $m$ and $n$ inside the small circles of the same gate signify that the $(mn)$-th block of the diagonal matrix CCD is of the form $diag(1,e^{\frac{i2\pi}{d^{k}}mn},e^{\frac{i2\pi}{d^{k}}2mn}, \ldots ,    e^{\frac{i2\pi}{d^{k}}(d-1)mn})$ while the rest of the blocks are identity matrices (see Eq. \eqref{eq:CCD}). That is the gate transforms the target qudit with the matrix $diag(1,e^{\frac{i2\pi}{d^{k}}mn},e^{\frac{i2\pi}{d^{k}}2mn}, \ldots ,    e^{\frac{i2\pi}{d^{k}}(d-1)mn})$ iff the control states are $|x\rangle = |m\rangle$ and $|y\rangle = |n\rangle$.

The way to construct a double controlled rotation gate $CCD_{(m,n)}(\varphi_{1},\varphi_{2},\ldots,\varphi_{d-1})$ is analogous to the one for the  simply controlled gate  $CD_{(m)}(\varphi_{1},\varphi_{2},\ldots,\varphi_{d-1})$ which is equivalent to the $CD^{\prime}_{m}(a_{1},a_{2},\ldots,a_{d-1})$. The difference in this case is that  we need  double controlled generalized NOT gates, which will be called $GCCX_{(m,n)}^{(jk)}$. They can be thought as an extension of Toffoli gates to the qudit case and their operation is analogous to that of  the $GCX_{(m)}^{(jk)}$, but in three qudits. They are defined by the equation 

\begin{equation}\label{eq:GCCX1}
\begin{array}{llr}
GCCX_{(m,n)}^{(jk)}= & |m \rangle \langle m| \otimes 
|n \rangle \langle n| \otimes
\left( |j \rangle \langle k| + |k \rangle \langle j| +
 \sum\limits_{
 \begin{subarray}{l}
 k=0 \\
 k \neq j 
 \end{subarray}}^{d-1} |k \rangle \langle k| \right) + & \\
 &
 \sum\limits_{
 \begin{subarray}{l}
 l=0 \\
 l \neq m \\
 l \neq n
 \end{subarray}}^{d-1} |l \rangle \langle l| \otimes I_{d} \otimes I_{d}

& j,k,m,n=0 \ldots d-1
\end{array}
\end{equation}
\noindent

\begin{figure}[!b]
\centerline{\epsfig{file=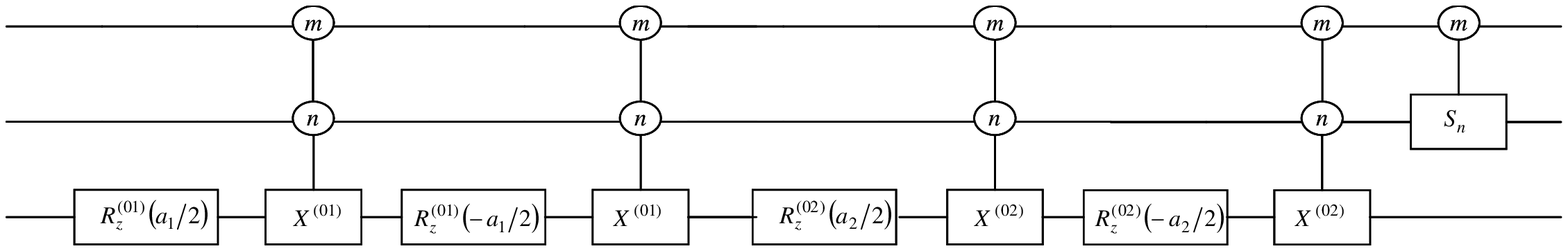, width=14cm}} 
\vspace*{13pt}
\fcaption{\label{fig:CCDDecomp}Decomposition of $CCD_{(m,n)}^{\prime}(a_{1},a_{2},\ldots,a_{d-1})$  gate}
\end{figure}

This description means that they interchange the two target qudit basis states $|j\rangle$ and $|k\rangle$ iff the two control states are $|m\rangle$ and $|n\rangle$. Having available the $GCCX_{(m,n)}^{(jk)}$ gates we can construct a $CCD^{\prime}_{(m,n)}(a_{1},a_{2},\ldots,a_{d-1})$ as shown in Figure \ref{fig:CCDDecomp} (the example depicts the qutrit case of $d=3$, generalization for other values of $d$ is obvious). The controlled gate $S_{n}$ is the analogous of the single qudit gate $S_{m}$ of Figure \ref{fig:CDgate}. Namely, it is a $CD_{m}(1,\ldots,1,\underset{\footnotesize n\textnormal{-th}\, \textnormal{pos}}{e^{i\varphi}},1,\ldots,1)$ gate, with angle $\varphi=\frac{1}{d}\sum_{k=1}^{d-1}a_{k}$. The $CCD_{(m,n)}^{\prime}$ gate is identical with the desired $CCD_{(m,n)}$ gate if angles redefinition similar to the ones of Eq. \eqref{eq:DDprime} are applied.


\begin{figure}[!t]
\centerline{\epsfig{file=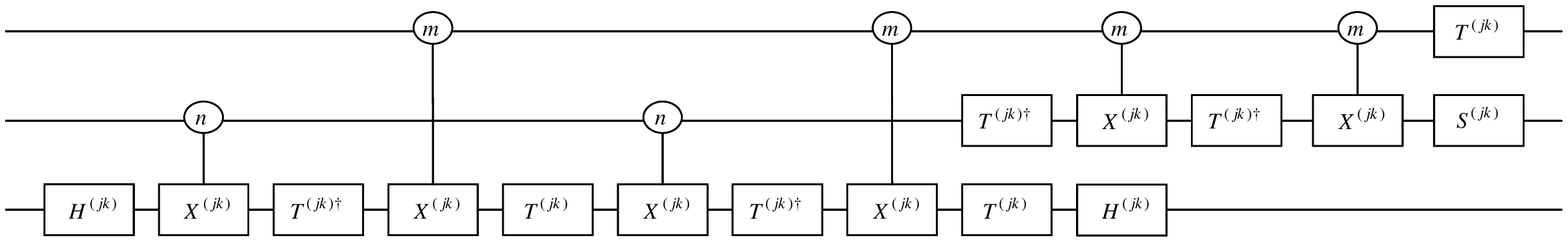, width=14cm}} 
\vspace*{13pt}
\fcaption{\label{fig:GCCDecomp}Decomposition of $GCCX_{(m,n)}^{(jk)}$  gate into single and two qudit two-level gates.}
\end{figure}

What remains is the $GCCX_{(m,n)}^{(jk)}$ gate  construction. This gate operates in a two dimensional subspace of the target qudit. Thus, extension to the qudit case of a Toffoli gate decomposition into single and two qubits \cite{Nie:2010} can be exploited and this is shown in Figure \ref{fig:GCCDecomp}. The single and two qudit gates used will be the generalization of the $S$,$T$ and $H$ qubit gates to the $d$ dimension of the qudits but operating only on a 2-dimensional subspace. Concretely, we define the gates 

\begin{equation}\label{eq:Sjk}
\begin{array}{lr}
S^{(jk)}=|j \rangle \langle j| + i|k \rangle \langle k| +
\sum\limits_{\begin{subarray}{l}
n=0 \\
n \neq j \\
n \neq k
\end{subarray}}^{d-1} |n \rangle \langle n|
&
j,k=0 \ldots d-1
\end{array}
\end{equation}
\noindent  

\begin{equation}\label{eq:Tjk}
\begin{array}{lr}
T^{(jk)}=|j \rangle \langle j| + e^{i\pi /4}|k \rangle \langle k| +
\sum\limits_{\begin{subarray}{l}
n=0 \\
n \neq j \\
n \neq k
\end{subarray}}^{d-1} |n \rangle \langle n|
&
j,k=0 \ldots d-1
\end{array}
\end{equation}
\noindent  

\begin{equation}\label{eq:Hjk}
\begin{array}{lr}
H^{(jk)}=\frac{1}{\sqrt{2}} \left( |j \rangle \langle j| + |j \rangle \langle k| + 
|k \rangle \langle j| - |k \rangle \langle k| \right) +
\sum\limits_{\begin{subarray}{l}
n=0 \\
n \neq j \\
n \neq k
\end{subarray}}^{d-1} |n \rangle \langle n|
&
j,k=0 \ldots d-1
\end{array}
\end{equation}
\noindent

Gates $S^{(jk)}$ and $T^{(jk)}$ are effectively $R_{z}^{(jk)}(\theta)$ with $\theta$ equal to $\pi /2$ and $\pi /4$, respectively, ignoring some global phase. Also, we can build the $H^{(jk)}$ gate using the easily proved identity $H^{(jk)}=e^{i\pi /2} R_{z}^{(jk)}(\pi /2) R_{x}^{(jk)}(\pi /2) R_{z}^{(jk)}(\pi /2)$, which is similar to the one for the qubit Hadamard case.
We have finally achieved to synthesize a three qudits rotation gate $\overline{R} _{k}^{(d)}$ with elementary single qudit rotation $R_{z}^{(jk)}(\theta)$, $R_{x }^{(jk)}(\theta)$ gates and two qudits $GCX_{(m)}^{(jk)}$ gates.

\appendix

\noindent 
The design of arithmetic quantum circuits based on the QFT involves a library of elementary gates  $\{ H^{(d)},  R_{z}^{(jk)}(\theta), R_{x}^{(jk)}(\theta), GCX_{m}^{(jk)}  \}$. The size of this library is not constant as the parameter $\theta$ of the rotation gates depends on the size $q$ of the circuit (number of qudits used). However, it is possible to approximate these rotation gates with arbitrary precision using a constant set of gates.  Much research has been done recently in this area focused on gates operating on qubits \cite{Har:2002,Fow:2011,Cod:2012,Boc:2012,Pha:2013,Duc:2013,Kli:2013a,Boc:2013,Kli:2013b,Sel:2015,Ros:2015,Kli:2016}. An extension of some of these results can be easily applied to the case of the qudits for the specific elementary gates library and thus we can use a constant library for the synthesis. This is important both for physical implementation reasons and for the fault tolerance aspect of the circuit, as fault tolerance techniques have developed for a restricted set of gates (e.g Clifford + T gates for the case of binary qubits). 
Before proceeding to the extension of the well established qubit gates approximation methods to the qudit case, some definitions are necessary. A unitary matrix $U$ of dimensions $d \times d$ is called a two level matrix if it has the form  \cite{Nie:2010}

\begin{equation}\label{eq:U2}
U=\begin{bmatrix}
1 	& 		& 		& 			&	 	& 		& 		& 		\\
	& \ddots & 		& 			& 		& 		& 		& 		\\
  	&        & 1 	& 			& 		& 		& 		& 		\\
  	&        &  	& u_{jj}		& 		& 	u_{jk}	& 		& 		\\
  	&        &  	& 		& 	1	& 		& 		& 		\\
  	&        &  	& u_{kj}		& 		& 	u_{kk}	& 		& 		\\
  	&  & 	& 			& 		& 			& \ddots		& 		\\	
 	& 		& 		& 			&	 	& 		& 		& 	1	\\  	
\end{bmatrix}
\end{equation}
\noindent

This kind of matrix leaves invariant a subspace of $d-2$ dimensions and operates only on the two dimensions corresponding to coordinates $j$ and $k$. A more compact notation for the above matrix is 

\begin{equation}\label{eq:U2compact}
\begin{array}{cc}
U_{\left[ jk \right]}(V), & 
V=
\begin{bmatrix}
v_{11} & v_{12} \\
v_{21} & v_{22}
\end{bmatrix}
\end{array}
\end{equation}
\noindent

In this notation, we only need to define a unitary $2 \times 2$ dimensional matrix $V$ and also define which coordinates $j$ and $k$ this matrix operates on. A multiplication homomorphism is valid since

\begin{equation}\label{eq:homo}
U_{\left[ jk \right]}(V_{1})\cdot 
U_{\left[ jk \right]}(V_{2}) =
U_{\left[ jk \right]}(V_{1} \cdot V_{2})
\end{equation}
\noindent

The $R_{z}^{(jk)}(\theta)$ and $R_{x}^{(jk)}(\theta)$ elementary gates have exactly the form of Eq. \eqref{eq:U2compact} with $V_{z}(\theta)=\begin{bmatrix}
e^{- i\theta/2} & 0 \\
0 & e^{ i\theta/2}
\end{bmatrix}$
and 
$V_{x}(\theta)=\begin{bmatrix}
\cos{\theta /2} & -i\sin{\theta /2} \\
-i\sin{\theta /2} & \cos{\theta /2} 
\end{bmatrix}$
, respectively.  The $V_{z}(\theta)$ and $V_{x}(\theta)$ gates are in fact qubits rotation gates  and thus we can  exploit the known approximation results of the literature for the qubit gates. These results state that an arbitrary rotation gate like $V_{z}(\theta)$ can be approximated by a finite sequence of gates belonging to a discrete set, e.g.  $\hat{V}_{z}(\theta)=(HT\cdots T)(HT\cdots T)\cdots (HT\cdots T)$, where $H$ is the Hadamard gate and $T$ is the $\pi/8$ gate, such as the approximation error $\epsilon = \| \hat{V}_{z}(\theta) - V_{z}(\theta) \|$ can be arbitrary small (Solovay-Kitaev Theorem and improvements \cite{Kit:1997,Kit:1999,Daw:2006}). Using this fact and Eq. \eqref{eq:homo} we find that every rotation gate $R_{z}^{(jk)}(\theta)$ can be approximated by another one $\hat{R}_{z}^{(jk)}(\theta)$ with arbitrary precision as 

\begin{equation}\label{eq:Rzapprox}
\hat{R}_{z}^{(jk)}(\theta)=
U_{\left[jk \right]}(\hat{V}_{z}(\theta) )=
U_{\left[jk \right]}(H) U_{\left[jk \right]}(T) \cdots 
U_{\left[jk \right]}(T) 
\cdots
U_{\left[jk \right]}(H) U_{\left[jk \right]}(T) \cdots 
U_{\left[jk \right]}(T)
\end{equation}
\noindent

In essence, the $U_{\left[jk \right]}(H)$ and $U_{\left[jk \right]}(T)$ gates are the $H^{(jk)}$ and $T^{(jk)}$ gates of Eq. \eqref{eq:Hjk} and \eqref{eq:Tjk}, respectively.
On the other side, the $R_{x}^{(jk)}(\theta)$ gates can be decomposed using the identity $R_{x}^{(jk)}(\theta)=H^{(jk)} R_{z}^{(jk)}(\theta) H^{(jk)} $ and thus the proposed circuits can be synthesized using the discrete library of constant number of components shown in Table \ref{tab:lib}. The second column shows the number of different gates of the same family, which depends on the family parameters (none, $j$, $k$ and $m$). The constant library consists of a total of $(3+d)\frac{d(d-1)}{2}+1$ gates.

\begin{table}[!b]
\tcaption{Discrete library ef elementary gates.}
\begin{center}
\begin{tabular}{ll}
  \hline			
  Gate family & \# gates  \\
  \hline
 $H^{d}$ 	 	&   1			\\
 $H^{(jk)}$ 		& $d(d-1)/2$ 	\\
 $T^{(jk)}$ 		& $d(d-1)/2$ 	\\
 $S^{(jk)}$ 		& $d(d-1)/2$ 	\\
 $GCX_{m}^{(jk)}$	& $d^{2}(d-1)/2$ \\
  \hline  
  Total \# gates   & $(d+3)\frac{d(d-1)}{2}+1$ \\
  \hline
\end{tabular}
\end{center}
\label{tab:lib}
\end{table}

The first Solovay-Kitaev algorithms \cite{Kit:1997,Kit:1999,Daw:2006} generate a sequence of such gates of length $O(\log^{3.97}{(1/\epsilon)})$ and synthesis time in order of $O(\log^{2.71}{(1/\epsilon)})$. In the last few years extensive research resulted in great improvements both in terms of the sequence length and synthesis time. They used a diverse set of techniques (usage of ancilla or not, different libraries, approximate or exact synthesis etc). Some of the best results in terms of the generated sequence length can be found in \cite{Boc:2013,Sel:2015,Ros:2015,Kli:2016}. These works offer a length of less than $10\log{(1/\epsilon)}$  $T$ gates ($T$ gates are considered more costly if they are built fault-tolerantly). In the presented circuits, the worst case angle of a $R_{z}^{jk}(\theta)$  gate is $\theta=2\pi / d^{q}$, so the desired approximation error should be of the same order $\epsilon \approx 2\pi/d^{q}$. Consequently, each $R_{z}^{(jk)}(\theta)$ gate can be adequately approximated by a sequence of $H^{(jk)}$ and $T^{(jk)}$ gates of length of the order $10\log{(d^{q}/2\pi)} \approx 10q\log{d}$. Thus if we have to use a constant library of components due to implementation and/or fault tolerance reasons we have a linear  in $q$ multiplicative overhead in the quantum costs and depths calculated in Section \ref{sec:Complexity}.

\end{document}